\documentclass[floatfix, a4paper, amsfonts, amssymb, amsmath, reprint, showkeys, nofootinbib, twoside, superscriptaddress, aps, pra]{revtex4-2}

\usepackage{graphicx} 
\usepackage{physics}
\usepackage{mathtools}
\usepackage{nicefrac}
\usepackage{siunitx} 
\usepackage{color, colortbl} 
\usepackage[normalem]{ulem} 
\usepackage{float}
\usepackage{xr}
\usepackage{array}
\usepackage{hyperref}
\usepackage[T1]{fontenc}
\usepackage{notes2bib}
\usepackage{xurl}
\usepackage{anyfontsize} 

\bibnotesetup{
  note-name = ,
  use-sort-key = false
}

\newcommand{\pcm}[1]{\SI{#1}{\per\cm}}
\newcommand{\um}[1]{\SI{#1}{\micro\meter}}
\newcommand{\nm}[1]{\SI{#1}{\nano\meter}}
\newcommand{\us}[1]{\SI{#1}{\micro\second}}
\newcommand{\pct}[1]{\SI{#1}{\percent}}
\begin{document}
\title{Radiative local density of states in three-dimensional photonic\\band-gap crystals to interpret time-resolved emission} 

\author{Timon J. Vreman}

\author{Ad Lagendijk}

\author{Willem L. Vos}
\homepage{https://nano-cops.com}
\email{w.l.vos@utwente.nl}


\address{Complex Photonic Systems (COPS), University of Twente, 7500 AE Enschede, The Netherlands}


\begin{abstract}
We investigate the spontaneous emission of light in three-dimensional (3D) photonic crystals through theoretical calculations and simulations.
It is well known that spontaneous emission depends on the radiative local density of states (RLDOS).
Photonic band-gap crystals radically modulate the RLDOS, thereby controlling spontaneous emission.
We compare two different methods to calculate the RLDOS: the plane-wave expansion (PWE) method and the finite-difference time-domain (FDTD) method.
The PWE method directly calculates the RLDOS of an infinite photonic crystal, whereas the FDTD method simulates the RLDOS through the power emitted by a dipole in a finite photonic crystal.
We demonstrate that the methods yield similar frequency-dependent trends in the RLDOS, with relative differences of less than 12\% that originate from the different boundary conditions.
We employ the plane-wave expansion method to compute distributions of emission rates that are relevant to many optical experiments where quantum emitters are distributed within a crystal. 
Such distributions of emission rates enable us to compute and directly interpret the time-resolved decay as observed in experiments.
We expect that our results promote the RLDOS to the realm of optical design and products.
\end{abstract}

\maketitle

\section{Introduction}
Spontaneous emission is the fundamental process behind almost all light sources~\cite{Novotny2012book, hewitt2015book}, and is typically well described in the dipole approximation~\cite{Milonni1994bookQV}. 
Three-dimensional (3D) photonic band-gap crystals are eagerly pursued for their potential to inhibit or enhance the spontaneous emission~\cite{Bykov1972JETP, Yablonovitch1987PRL, noda2000science, Vats2002PRA, Leistikow2011PRL}.
Photonic crystals are periodic dielectric materials, where the periodicity is on the scale of the wavelength of light~\cite{Joannopoulos2008book}.
The rate of spontaneous emission, i.e., the radiative decay rate, of a rotating dipole located at a position $\rm{\mathbf{r}}_0$ emitting at a frequency $\omega$ is given by Fermi's golden rule~\cite{Fermi1932RMP} that is reformulated in nanophotonics as~\cite{Snoeks1995PRL, asatryan2001PRE, rockstuhl2009PRB, vos2009PRA, Novotny2012book, Lodahl2015RevModPhys, Nikolaev2009JOSAB}\bibnote{To use the RLDOS, the dipole should be rotating fast enough to average out directional variations in the---later specified---partial RLDOS.},
\begin{equation}\label{eq:ldos_FermiGoldenRule}
    \Gamma_\mathrm{rad}\left(\rm{\mathbf{r}}_0,\omega\right) = \frac{\pi\omega}{3\hbar \varepsilon_{\rm 0}}
    \left|\mathcal{P}\right|^2 N\left(\rm{\mathbf{r}}_0,\omega\right),
\end{equation}
where $\varepsilon_{\rm 0}$ is the permittivity of free space, $\mathcal{P}$ the transition dipole moment, and $N(\rm{\mathbf{r}}_0,\omega)$ the \textit{radiative} local density of states (RLDOS)~\cite{Sprik1996EPL}.
Photonic crystals enhance or inhibit spontaneous emission because their states, Bloch modes, are different from the modes in free space, thus modifying $N(\rm{\mathbf{r}}_0,\omega)$.

Multiple numerical methods are used to calculate the decay rate.
The plane wave expansion (PWE) method computes the RLDOS in an infinite photonic crystal without boundaries~\cite{Busch1998PRE,li2000PRL, wang2002PRL, Vats2002PRA, wang2003PRB, Nikolaev2009JOSAB}, obtaining the decay rate through Eq.~\ref{eq:ldos_FermiGoldenRule}.
The numerical finite-difference time-domain (FDTD) method simulates the emission of a dipole in a finite photonic crystal with boundaries~\cite{taflove2005book, hermann2002JOSAB, ishizaki2009JOSAB, Mavidis2020PRB}, obtaining the decay rate through the emitted power~\cite{xu2000pra, taflove2005book}.
Remarkably, to the best of our knowledge, these popular methods have not yet been compared to date, especially in the case of 3D photonic crystals.

Furthermore, current emission calculations in photonic crystals cannot be directly compared to many emission experiments.
Emission measurements often probe the time-resolved decay curve $f(t)$ of a large ensemble of emitters distributed at many positions in a 3D photonic crystal~\cite{petrov1998PRL, yoshino1998APL, megens1999commentPetrov, lodahl2004nature, Li2007AdvMat, ventura2008AdvMater, Nikolaev2007PRB, Vallee2007PRB, Vion2009JAP, Leistikow2011PRL, Jorgensen2011PRL, ning2012AdvMater, schulz2022acs}, where $f(t)$ is the number of detected photons per time bin in experiments.
The RLDOS has been calculated before in photonic crystals with various structures with the goal to interpret spontaneous emission experiments in photonic crystals~\cite{Busch1998PRE, li2000PRL, wang2002PRL, hermann2002JOSAB, wang2003PRB, Nikolaev2009JOSAB, ishizaki2009JOSAB, Mavidis2020PRB}.
However, most of these studies describe the emission by only one emitter at only one or a few positions inside a 3D photonic crystal~\cite{Busch1998PRE, li2000PRL, hermann2002JOSAB, wang2003PRB, Nikolaev2009JOSAB, Mavidis2020PRB}.
While Wang \textit{et al.} do consider multiple emitter positions~\cite{wang2002PRL}, calculating $f(t)$ from their calculated distributions of lifetimes is difficult~\bibnote{While the relation between decay rate and lifetime is simple for a single emitter, the relation between distributions of lifetimes and decay rates of an ensemble of emitters that have varying decay rates is complex. Therefore, Eq.~\ref{eq:ldos_ft_to_zetaGamma} cannot directly be applied when using distributions of lifetimes $\zeta(\tau)$ instead of decay rates $\zeta(\Gamma)$.}.

Here, we perform new calculations on the RLDOS with the goal of interpreting spontaneous emission experiments. 
In the first part, we compute the RLDOS at a single position in (A) two different types of 3D photonic crystal structures, and (B) using the PWE and FDTD methods.
We demonstrate that the two methods yield RLDOSs in both photonic crystals of which the trends agree well, but also that the methods differ in details due to the presence or absence of finite-size effects.
In the second part, we calculate the RLDOS at many specific emitter positions using the PWE method, effectively in the same computational time as RLDOS calculations at a single position, similar to Wang \textit{et al.}~\cite{wang2002PRL}. 
We use this ensemble of RLDOSs relative to a reference RLDOS to calculate the decay curve of the whole ensemble of emitters present at these many positions.
The decay curve is directly obtained in previously mentioned experiments via time-correlated single-photon counting measurements~\cite{Lakowicz2006book}.

\section{Methods}
\subsection{Structures}
\begin{figure}[t]
    \centering
    \includegraphics[width=\linewidth]{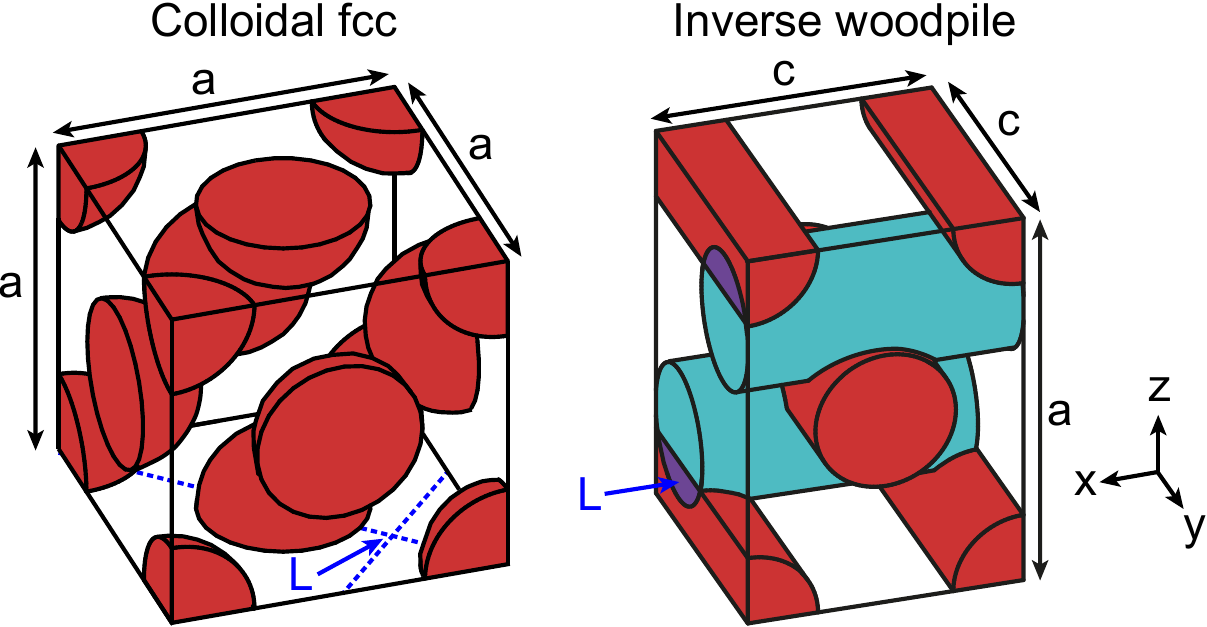}
    \caption{Illustrations of non-primitive unit cells. 
    (Left) The spheres of the fcc colloidal opal with pitch $a$.
    (Right) The pores of the inverse-woodpile structure with pitches $a$ and $c \equiv a/\sqrt{2}$.
    The RLDOS locations, due to symmetry equivalent to those in the text, are denoted by $L$.}
    \label{fig:ldos_structures}
\end{figure}

We study two different 3D photonic crystal structures to verify the robustness of our results: a face-centered cubic (fcc) colloidal crystal without a band gap, and an inverse-woodpile crystal with a 3D band gap, as shown in Fig.~\ref{fig:ldos_structures}.
The fcc colloidal crystal has been studied often to test local density of states calculations~\cite{Busch1998PRE, wang2003PRB, Nikolaev2009JOSAB}. 
Using the fcc unit cell from Ref.~\cite{Joannopoulos2008book}, the fcc lattice has a cubic lattice parameter $a =$ \nm{680}, with spheres with $\varepsilon =$ 7.35 (typical for TiO$_2$, set constant as a function of $\omega$) and a sphere volume fraction of \pct{25}.
The sphere is positioned at $(0,0,0)$ inside the unit cell and is repeated by the primitive lattice vectors.
The interstitial space is filled with a dielectric medium with $\varepsilon =$ 1.77, typical of water.

The second photonic crystal we investigate features the cubic diamond-like inverse woodpile structure with a complete photonic band gap~\cite{Ho1994SSC}, as studied in various experiments~\cite{Hillebrand2003JAP, Leistikow2011PRL, schulz2022acs, vreman2025thesis}\bibnote{Effectively, the inverse woodpile structure also has the fcc structure like the colloidal crystal, albeit with a much more complex crystal basis~\cite{Ashcroft1976book}.}.
The structure contains pores of $n = 1.48$ ($\varepsilon\approx 2.19$ as is typical for toluene, diameter $d_{\rm{pore}} =$ \nm{245}, pitch $a =$ \nm{680}) with a backbone of $\varepsilon = 12.1$ (typical for silicon).
The pore along the $x$ direction is positioned at $(0, \frac{a}{4\sqrt{2}}, \frac{a}{4})$, along the $y$ direction at $(0,0,0)$, repeated by lattice vectors of the body-centered tetragonal lattice, see Ref.~\cite{vreman2025thesis}.

\subsection{Finite-difference time-domain method}
With the FDTD method, we simulate a dipole emitter at position $\vb{r}_0$ with a dipole moment oriented in direction $\vb{e}_d$ to calculate the emitted power from which the \textit{partial} RLDOS, $N_P(\vb{r}_0, \omega, \vb{e}_d)$, is derived. 
The partial RLDOS is equal to the RLDOS in the direction of the dipole moment $\vb{e}_d$~\cite{Novotny2012book}, and is given by~\cite{taflove2005book}
\begin{equation}
    N_P(\vb{r}_0, \omega, \vb{e}_d) = \frac{4}{\pi}P(\vb{r}_0,\omega, \vb{e}_d, \vb{e}_d),
\end{equation}
where $P(\vb{r}_0,\omega, \vb{e}_d)$ is the emitted power of the dipole.
To calculate the RLDOS, we use the open-source FDTD software package MEEP~\cite{oskooi2010elsevier}.
The FDTD method time steps the electromagnetic fields in real space, $(t, \vb{r})$.
The fields are absorbed by a perfectly matched layer of \um{1.4} around the photonic crystal, mimicking free space.
The output power is recorded and the partial RLDOS is calculated by internal functions in MEEP.
The RLDOS is found by summing the partial RLDOS obtained from simulations along three orthogonal directions ($\vb{e}_x$, $\vb{e}_y$, $\vb{e}_z$)~\cite{vos2009PRA}, i.e., $N(\vb{r}_0, \omega) = N_{P}(\vb{r}_0, \omega, \vb{e}_x) + N_{P}(\vb{r}_0, \omega, \vb{e}_y) + N_{P}(\vb{r}_0, \omega, \vb{e}_z)$.

Similar to experiments, the crystals under study are finite; thus, crystal surfaces are intrinsically present.
While unavoidable defects may be considered, this is not typically done as it places a large burden on computational efforts. 

\subsection{Plane-wave expansion method}
The PWE method computes the modes and frequencies of a photonic crystal in the frequency domain by expanding the periodic dielectric function (the scattering potential) and the Bloch modes in a sum of reciprocal lattice vectors in wave vector space $(\omega, \vb{k})$.
Inherently, the PWE method considers an infinite photonic crystal.
Thus, neither external crystal surfaces are considered, nor any unavoidable deviations from perfect periodicity, such as point defects.

We use the open-source PWE package MIT Photonic Bands~\cite{johnson2001OptExp} to calculate the electric fields $\rm{\mathbf{E}}_{\it{i}}\left(\rm{\mathbf{r}}_0, \omega_{\it i}\right)$ and frequencies $\omega_i$ of the Bloch modes in the photonic crystals.
Due to the periodicity, all modes are obtained by sampling the modes at $\vb{k}$-vectors in the top half of the first Brillouin zone~\cite{wang2003PRB, Nikolaev2009JOSAB}.
The RLDOS is calculated by a weighted counting of the available modes~\cite{Snoeks1995PRL, Busch1998PRE, Nikolaev2006thesis, Novotny2012book}:
\begin{equation}\label{eq:ldos_ldos}
    N\left(\rm{\mathbf{r}_0},\omega\right) = \sum_i\delta\left(\omega-\omega_i\right)\left|\rm{\mathbf{E}}_{\it{i}}\left(\rm{\mathbf{r}}_0, \omega_{\it i}\right)\right|^2,
\end{equation}
where the weights are the electric fields normalized according to
\begin{equation}\label{eq:ldos_Enormalization}
\int_V{\varepsilon(\vb{r})\rm{\mathbf{E}}_{\it{i}}\left(\rm{\mathbf{r}}, \omega_i\right)\cdot \rm{\mathbf{E}}_{\it{j}}\left(\rm{\mathbf{r}}, \omega_j\right)} d^3 \rm{\mathbf{r}} = \delta_{\it{i}\rm{,}\it{j}},
\end{equation}
where $\delta_{i,j}$ is the Kronecker delta function and $V$ the volume of the unit cell.


\subsection{RLDOS at many positions}{\label{sec:ldos_rldos_manyPos}}
So far, we have considered the emission of a single emitter at a single position $\vb{r}_0$.
However, multiple emitters are often probed simultaneously in experiments, and these emitters may be distributed inhomogeneously throughout the crystal at positions $\{\vb{r}\}$ within the unit cell. 
The PWE method allows us to calculate the RLDOS at these positions, $N(\{\vb{r}\}, \omega)$, using Eq.~\ref{eq:ldos_ldos}.
The calculation takes hardly any extra time because $\vb{E}_i(\vb{r}, \omega_i)$ is already obtained when calculating the RLDOS at $\vb{r}_0$. 
The RLDOS at each $\vb{r}$ is converted to a decay rate through Eq.~\ref{eq:ldos_FermiGoldenRule}.

\subsection{Analytical reference RLDOS in a homogeneous medium}{\label{sec:ldos_homRLDOS}}
The RLDOS in a homogeneous medium does not depend on position and can be calculated analytically.
Therefore, homogeneous media serve as an excellent reference.
The radiative local density of states in a homogeneous medium with refractive index $n$ is equal to~\cite{loudon2000book, Novotny2012book}
\begin{equation}
    N_{\rm{hom},\it{n}}(\omega) \,\rm{d}\omega = \frac{\it{n}}{\pi^2 \it{c}_{\rm{0}^{\rm{3}}}}\omega^2 \,\rm{d}\omega,
\end{equation}
where $\omega$ is the angular frequency and $c_0$ the speed of light in vacuum.
We rewrite the RLDOS as a function of various frequency measures as
\begin{align}
     &N_{\rm{hom},\it{n}}\left(\frac{a}{\lambda}\right) \,\rm{d}\omega = \left(\frac{4\it{n}}{\it{a}^{\rm{2}} c_{\rm{0}}}\right) \frac{\it{a}^{\rm{2}}}{\lambda^2}\, \rm{d}\omega  \label{eq:ldos_ldosADivLambda1}\\
      &N_{\rm{hom},\it{n}}\left(\frac{a}{\lambda}\right) \,\rm{d}\!\left(\frac{\it{a}}{\lambda}\right) = \left(\frac{8\pi \it{n}}{\it{a}^{\rm{2}} }\right) \frac{\it{a}^{\rm{2}}}{\lambda^2}\,\rm{d}\!\left(\frac{\it{a}}{\lambda}\right) \label{eq:ldos_ldosADivLambda2} \\[3pt]
     &N_{\rm{hom},\it{n}}(\tilde{\nu}) \,\rm{d}\tilde{\nu} = 8\pi \it{n}  \tilde{\nu}^{\rm{2}}\, \rm{d}\tilde{\nu},\label{eq:ldos_ldosHomWnr}
\end{align}
where $a$ is the (fictitious) pitch, $\lambda$ the wavelength of light in vacuum, and $\tilde{\nu} = \frac{\omega}{2\pi c_0} = \frac{1}{\lambda}$ the wavenumber.
The local density of states is sometimes reported in arbitrary units, see e.g. Refs.~\cite{Sprik1996EPL, Busch1998PRE, wang2002PRL}, which unfortunately does not provide much information to use or reproduce the data.
In other cases, e.g. Ref.~\cite{Nikolaev2009JOSAB}, Eq.~\ref{eq:ldos_ldosADivLambda1} is used, which is not practical because it mixes two different frequency scales.
Equation~\ref{eq:ldos_ldosADivLambda2} gives the RLDOS in purely reduced frequencies, which is convenient for theoretical studies.
We prefer the experimental scale wavenumber $\tilde{\nu}$~\cite{wikiWavenumber}, which yields the simple Eq.~\ref{eq:ldos_ldosHomWnr} for the RLDOS with units \SI{}{\centi\meter^{-2}}.
Both computational methods in this paper have been tested with various homogeneous media, and they always agree very well with Eq.~\ref{eq:ldos_ldosHomWnr}.

Because the RLDOS does not depend on position, there is just a single decay rate $\Gamma_{\rm{ref}}$ according to Eq.~\ref{eq:ldos_FermiGoldenRule}---assuming all emitters are the same. 
The single-exponential reference decay curve is equal to
\begin{equation}{\label{eq:referenceCurve}}
    f_{\rm{ref}}(t) = A_{\rm{ref}}\Gamma_{\rm{ref}}e^{-\Gamma_{\rm{ref}} t},
\end{equation} 
where $A_{\rm{ref}}$ is the amplitude, experimentally the number of counts at arrival time $t = 0$.

\subsection{RLDOS distributions to calculate the decay curve}
Experiments probe the decay curve $f(t)$ through time-correlated single-photon counting, where $f(t)$ is a histogram of photon arrival times that are collected upon repeated excitation in a time-correlated single-photon counting experiment~\cite{Lakowicz2006book}.
The decay curve of emitters in the photonic crystal $f_{\rm{phc}}(t) \equiv f(t)$ at a wavenumber $\tilde{\nu}_c$ is compared to a reference decay curve $f_{\rm{ref}}(t)$ at $\tilde{\nu}_c$ to calculate the enhancement.
Our computational approach is similar: 
We calculate the decay curve $f(t)$ at $\tilde{\nu}_c$ of many emitters inhomogeneously distributed at positions $\{\vb{r}\}$ in a photonic crystal, relative to a decay curve in a homogeneous reference medium $f_{\rm{ref}}(t)$ at $\tilde{\nu}_c$.

From calculations, we know the RLDOSs at the positions of the emitters $\{\vb{r}\}$ in the photonic crystal $N_{\rm{phc}}(\{\vb{r}\}, \tilde{\nu}) \equiv N(\{\vb{r}\}, \tilde{\nu})$ (Sec.~\ref{sec:ldos_rldos_manyPos}) and the RLDOS in the reference medium $N_{\rm{ref}}(\tilde{\nu})$ (Sec.~\ref{sec:ldos_homRLDOS}).
The relative RLDOSs are obtained by dividing $N(\{\vb{r}\}, \tilde{\nu})$ by $N_{\rm{ref}}(\tilde{\nu})$.
We construct a normalized histogram of the relative RLDOSs in a narrow wavenumber regime $\Delta\tilde{\nu}$ centered around $\tilde{\nu}_c$: 
\begin{equation}
    \rho\left(\frac{N(\{\vb{r}\}, \tilde{\nu}_c \pm (\Delta\tilde{\nu}/2))}{N_{\rm{ref}}(\tilde{\nu}_c \pm (\Delta\tilde{\nu}/2))}\right) \equiv \rho\left(\frac{N}{N_{\rm{ref}}}\right).
\end{equation}
For simplicity assuming all emitters are equal and have a quantum efficiency of \pct{100}, $\rho\left(\frac{N}{N_{\rm{ref}}}\right)$ is equal to the distribution of relative decay rates $\rho\left(\frac{\Gamma}{\Gamma_{\rm{ref}}}\right)$ using Eq.~\ref{eq:ldos_FermiGoldenRule}.
Using the fact that $\Gamma_{\rm{ref}}$ is known from the decay curve $f_{\rm{ref}}(t)$ (see Eq.~\ref{eq:referenceCurve}), we calculate the distribution of decay rates in the photonic crystal, $\rho(\Gamma)$.
Finally, we compute the distribution 
\begin{equation}
\zeta(\Gamma) = A\Gamma\rho(\Gamma), 
\end{equation}
which relates to the decay curve $f(t)$ via the Laplace transform~\cite{vanDriel2007PRB}
\begin{equation}\label{eq:ldos_ft_to_zetaGamma}
    f(t) = \int^\infty_0\zeta(\Gamma)\exp{-\Gamma t}\,\rm{d}\Gamma.
\end{equation}
When the number of excited emitters is the same in the photonic crystal as in the reference, the amplitude $A$ equals $A_{\rm{ref}}$.

\section{Results}
\subsection{Colloidal crystal at a single position}
\begin{figure}[t]
    \centering 
    \includegraphics[width=0.95\linewidth]{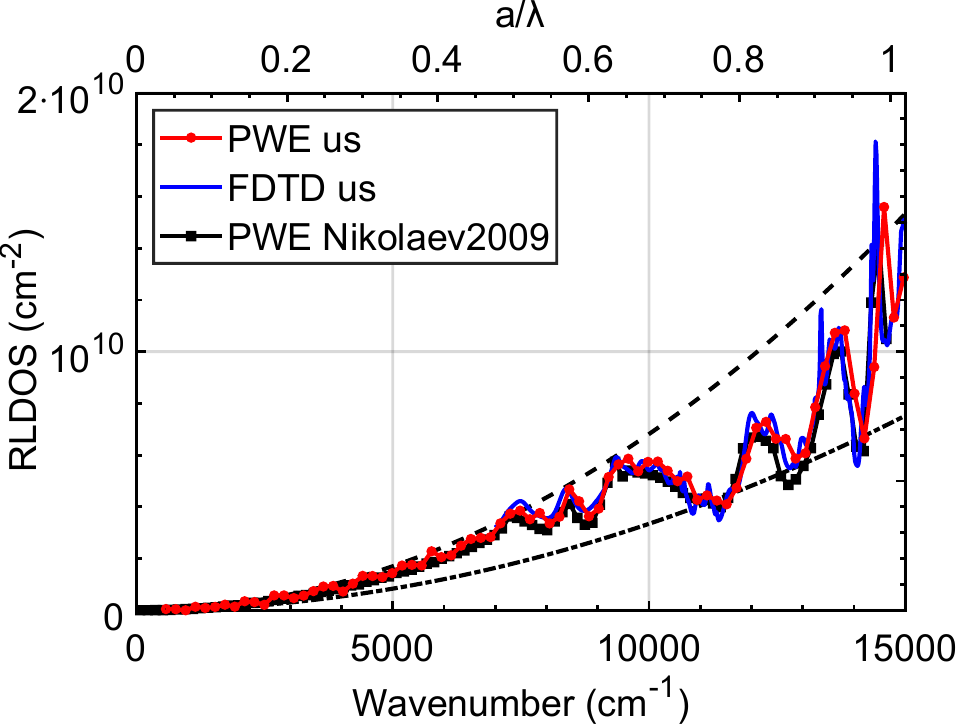}
    \caption{
    Radiative local density of states (RLDOS) calculated at $(\frac{a}{4},\frac{a}{4},0)$ inside the colloidal photonic crystal using the (red) PWE and (blue) FDTD method.
    (Black solid line) Calculations by Nikolaev \textit{et al.} using the PWE method~\cite{Nikolaev2009JOSAB}.
    (Black dashed lines) RLDOS of homogeneous media, namely TiO$_2$ (dashed, $\varepsilon = 7.35$) and water (dash-dotted, $\varepsilon = 1.77$).
    }
    \label{fig:ldos_res_fcc}
\end{figure}

We calculate the radiative local density of states (RLDOS) inside the fcc colloidal crystal at position $(\frac{a}{4},\frac{a}{4},0)$ using two different methods.
We first turn to the plane wave expansion method in Fig.~\ref{fig:ldos_res_fcc} (red).
For this calculation, we use \num{1728} plane waves, and \num{3456} $\vb{k}$-points in the top half of the first Brillouin zone~\cite{wang2003PRB}. 
These and the following quantities are chosen as a trade-off between accuracy and computation time. 
The results of PWE calculations by Nikolaev \textit{et al.}~\cite{Nikolaev2009JOSAB} (black) are also shown, who use more $\vb{k}$-points (\num{59520}) but fewer plane waves (\num{965}). 
The peaks and troughs in the RLDOS match closely, giving an error $L_1(y_1, y_2) \equiv \texttt{mean}(\abs{y_2-y_1})$ of $L_1 = $ \SI{4e8}{\centi\meter^{-2}}, which is 9$\!\,\times$ smaller than the mean RLDOS.
As discussed by Nikolaev \textit{et al.}, who compare their results to Ref.~\cite{wang2003PRB}, small differences are expected, especially given the differences in the number of $\vb{k}$-points and plane waves.
Therefore, we conclude that our PWE method agrees well with the literature.

In addition, we simulate the RLDOS at the same position as the PWE method within a $6a\times6a\times6a$ colloidal photonic crystal using the FDTD method with a resolution of 25 pixels per $a$.
The FDTD method exhibits a few sharp peaks, which we attribute to the finite size of the sample and the presence of surfaces.
The trends of the RLDOS obtained by the FDTD and PWE method agree well with a relatively small error of $L_1 = $ \SI{7e8}{\centi\meter^{-2}}.
Therefore, our calculations show that determining the RLDOS through the FDTD or PWE method is nearly equivalent in the studied $6a\times6a\times6a$ colloidal photonic crystal.

We also calculate the RLDOS of two homogeneous media using Eq.~\ref{eq:ldos_ldosHomWnr} in Fig.~\ref{fig:ldos_res_fcc} (black dashed), namely water ($\varepsilon_{\rm{low}} = 1.77$) and TiO$_2$ ($\varepsilon_{\rm{high}} = 7.35$), the materials from which the photonic crystal is made.
At low wavenumbers, $\tilde{\nu} <$ \pcm{7000} ($\frac{a}{\lambda} < 0.48$), the increase in the RLDOS of the photonic crystal is parabolic, similar to that of a homogeneous medium.
Fitting Eq.~\ref{eq:ldos_ldosHomWnr}, we find that the crystal behaves as an effective homogeneous medium for the RLDOS with $\varepsilon = 6.1$ up to $\tilde{\nu} =$ \pcm{7000}.
It is remarkable that the effective $\varepsilon$ of the homogeneous medium is closer to $\varepsilon_{\rm{high}}$ whereas $\vb{r}_0$ is located in the $\varepsilon_{\rm{low}}$ medium; the reason for this is currently unknown.

We observe that at greater wavenumbers, the RLDOS in the photonic crystal oscillates between the low and high dielectric RLDOS, i.e., $N_{\rm{hom,water}}(\tilde{\nu}) \lesssim N_{\rm{phc}}(\vb{r}_0,\tilde{\nu}) \lesssim N_{\rm{hom,TiO_2}}(\tilde{\nu})$.
The RLDOS is converted to a decay rate of a fast-rotating emitter using Eq.~\ref{eq:ldos_FermiGoldenRule}.
We find that the decay rate of this emitter at $\vb{r}_0$ inside the photonic crystal compared to that in a reference medium with dielectric constant $\varepsilon_{\rm{ref}}$, $\Gamma/\Gamma_{\rm{ref}}$, will not be much greater than the enhancement of decay rate reached when the emitter is in a homogeneous medium of $\varepsilon_{\rm{high}}$ relative to the homogeneous reference medium. 
In other words, at $\vb{r}_0$, $\Gamma/\Gamma_{\rm{ref}} \lesssim \varepsilon_{\rm{high}}/\varepsilon_{\rm{ref}}$.
Conversely, the decay rate in this photonic crystal, relative to a reference medium with $\varepsilon_{\rm{ref}}$, often remains larger than the decay rate when the emitter is in a homogeneous medium of $\varepsilon_{\rm{low}}$, i.e., $\Gamma/\Gamma_{\rm{ref}} \gtrsim \varepsilon_{\rm{low}}/\varepsilon_{\rm{ref}}$.

\subsection{Inverse woodpile crystal at a single position}{\label{sec:ldos_iwp_rmiddle}}
\begin{figure}[t]
    \centering 
    \includegraphics[width=0.95\linewidth]{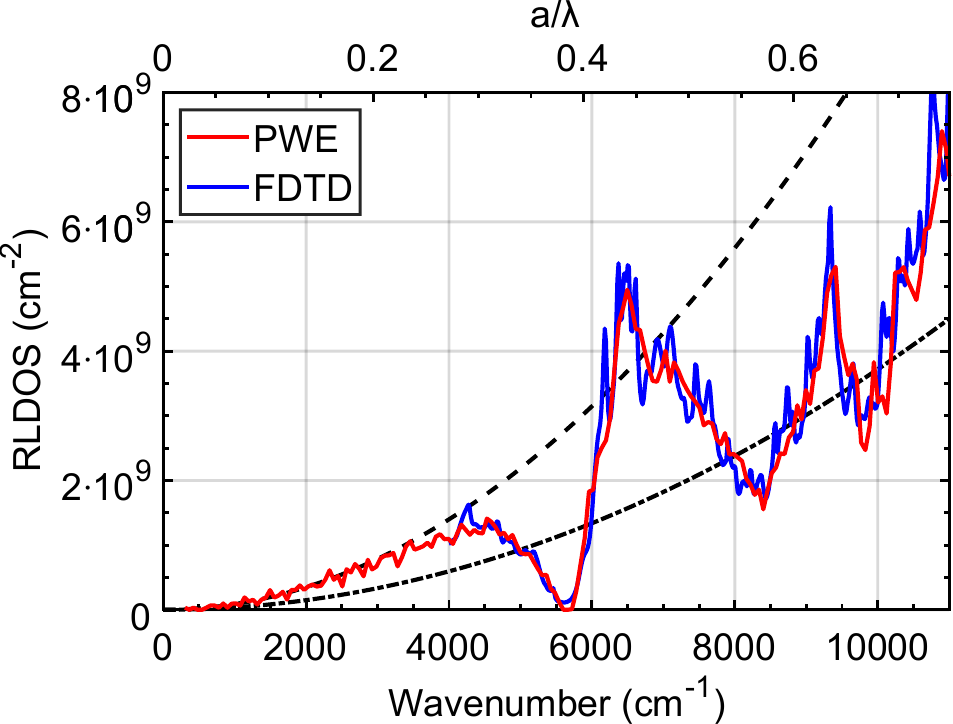}
    \caption{
    Radiative local density of states (RLDOS) calculated at (0,$\frac{c}{4}$,$\frac{a}{8}$) inside the inverse-woodpile photonic crystal using the (red) PWE and (blue) FDTD method.
    The pores of the silicon ($\varepsilon = 12.1$) crystal are filled with toluene ($n = 1.48$).
    (Black lines) RLDOS of the homogeneous media (dash-dotted) toluene, and (dashed) silicon.
    }
    \label{fig:ldos_res_iwp}
\end{figure}
We calculate the RLDOS in a 3D photonic band-gap crystal with an inverse woodpile structure at position $(0, \frac{a}{4\sqrt{2}}, \frac{a}{8})$, i.e., furthest away from silicon, at the center of the crossing between two pores.
For the PWE method, we use \num{6886} $\vb{k}$-points and \num{1728} plane waves.
The FDTD simulation simulates a structure of size $6c\times6c\times6a$, where $c\equiv a/\sqrt{2}$, using a grid with \num{27000} voxels per $a^3$.

The trends of the PWE calculation and FDTD simulation in Fig.~\ref{fig:ldos_res_iwp} resemble each other closely; the strong peaks and troughs obtained through the different methods nicely overlap, giving an error of $L_1 = $ \SI{4e8}{\centi\meter^{-2}}, or $L_1/\texttt{mean}(N(\vb{r}_0,\tilde{\nu})) <$ \pct{12}.
However, there appears to be speckle (`grass') modulated on top of the FDTD results.
The speckle becomes sharper as the resolution increases and the FDTD simulation runs longer.
Therefore, the speckle does not originate from numerical noise; instead, we attribute it to surface effects that are only present in the FDTD method: waves emitted by the dipole are reflected at the boundary of the photonic crystal.
Intuitively, reflections at the boundary may be stronger in an inverse structure---like the inverse woodpile---than in the direct colloidal structure defined above.
Other than the speckle, the most significant relative difference in RLDOS between the methods occurs in the photonic band gap, where the RLDOS is absolutely zero for the PWE method, but at a minimum $N(\tilde{\nu} =$ \pcm{5600}$) = $ \SI{1.1e8}{cm^{-2}} for the FDTD method. 
The RLDOS in the finite-size photonic crystal is 10 times lower than $N_{\rm{hom,toluene}}$ at the same wavenumber, which is consistent with Ref.~\cite{Mavidis2020PRB} because we simulate a larger photonic crystal.

Just like the RLDOS of the TiO$_2$ spheres in water, $N(\vb{r}_0,\tilde{\nu})$ at low wavenumbers is mainly determined by the high-$\varepsilon$ material, here silicon, while $\vb{r}_0$ is in the low-$\varepsilon$ material.
Here, we do observe that an enhancement higher than in homogeneous high-$\varepsilon$ material is possible, namely just above the band gap.
Apparently, the modes absent in the band gap are pushed towards higher wavenumbers.
Further above the band gap, just like at the probed position in the TiO$_2$ spheres in water, $N(\vb{r}_0,\tilde{\nu})$ remains lower than the homogeneous RLDOS $N_{\varepsilon,\rm{high}}(\tilde{\nu})$.

\subsection{Position-averaged RLDOS of the inverse woodpile photonic crystal}{\label{sec:ldos_iwp_pores}}

\begin{figure}[t]
    \centering 
    \includegraphics[width=0.95\linewidth]{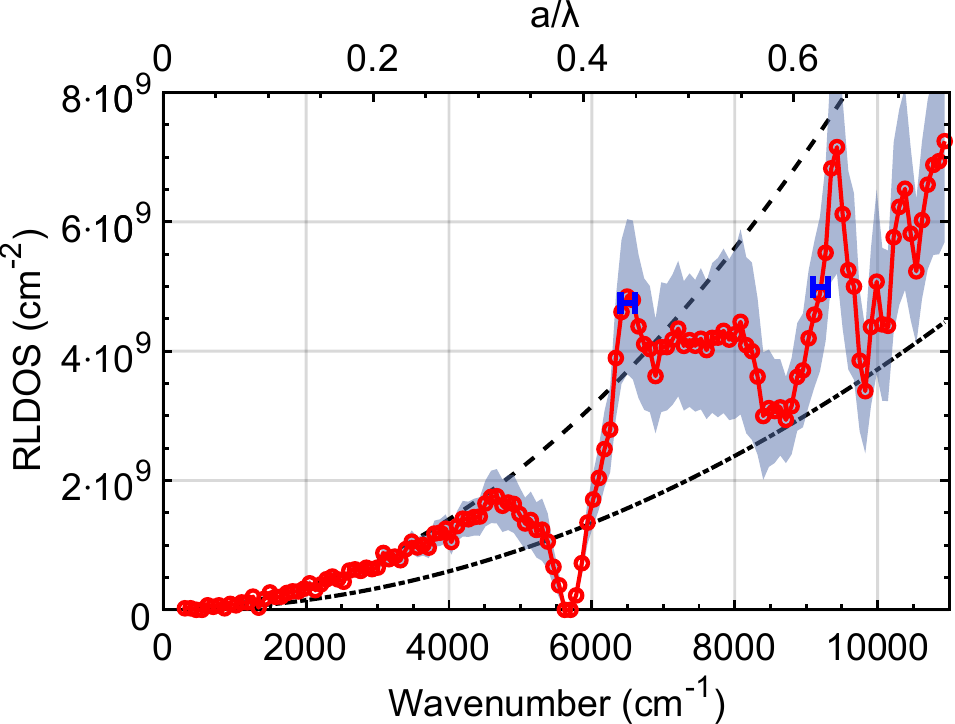}
    \caption{
    (Red) Radiative local density of states (RLDOS) averaged over all positions in the pores of the inverse-woodpile photonic crystal using the PWE method.
    (Blue shaded area) Standard deviation of the RLDOS around the average value.
    (Black dashed lines) LDOS of homogeneous media, namely toluene (the pores' material, $n = 1.48$) and silicon ($\varepsilon = 12.1$).
    (Blue bars) Regions in which we determine $\zeta(\Gamma)$ in Fig.~\ref{fig:ldos_zeta_ft_wnr6500}.
    }
    \label{fig:ldos_avg_iwp_pores}
\end{figure}
The 3D photonic band-gap crystal with the inverse woodpile structure consists of toluene-filled pores in silicon.
In experiments, quantum dot emitters are present at \textit{many} positions in the pores, while they cannot reside within the silicon backbone.
To calculate the average RLDOS in the pores, we use the same Bloch modes of the photonic crystal with an inverse-woodpile structure that we calculated in the previous section.
Of each of those computed Bloch modes, we extract the normalized electric field at \num{13824} positions, where the electric field grid is doubled in each dimension via interpolation.
At each position in the pores, we compute the RLDOS using Eq.~\ref{eq:ldos_ldos}.
The average RLDOS inside the pores of the inverse woodpile photonic crystal and its standard deviation are shown in Fig.~\ref{fig:ldos_zeta_ft_wnr6500}.
We discern that particular features in the RLDOS are not random but instead occur at many positions inside the pores, such as the peaks just after the photonic band gap and at $\tilde{\nu} =$ \pcm{9400}.
We also observe these features in the RLDOS calculated at a single position in Fig.~\ref{fig:ldos_res_iwp}.

\subsection{Decay curve in the pores of the inverse woodpile photonic crystal}

\begin{figure}[t]
    \centering 
    \includegraphics[width=0.85\linewidth]{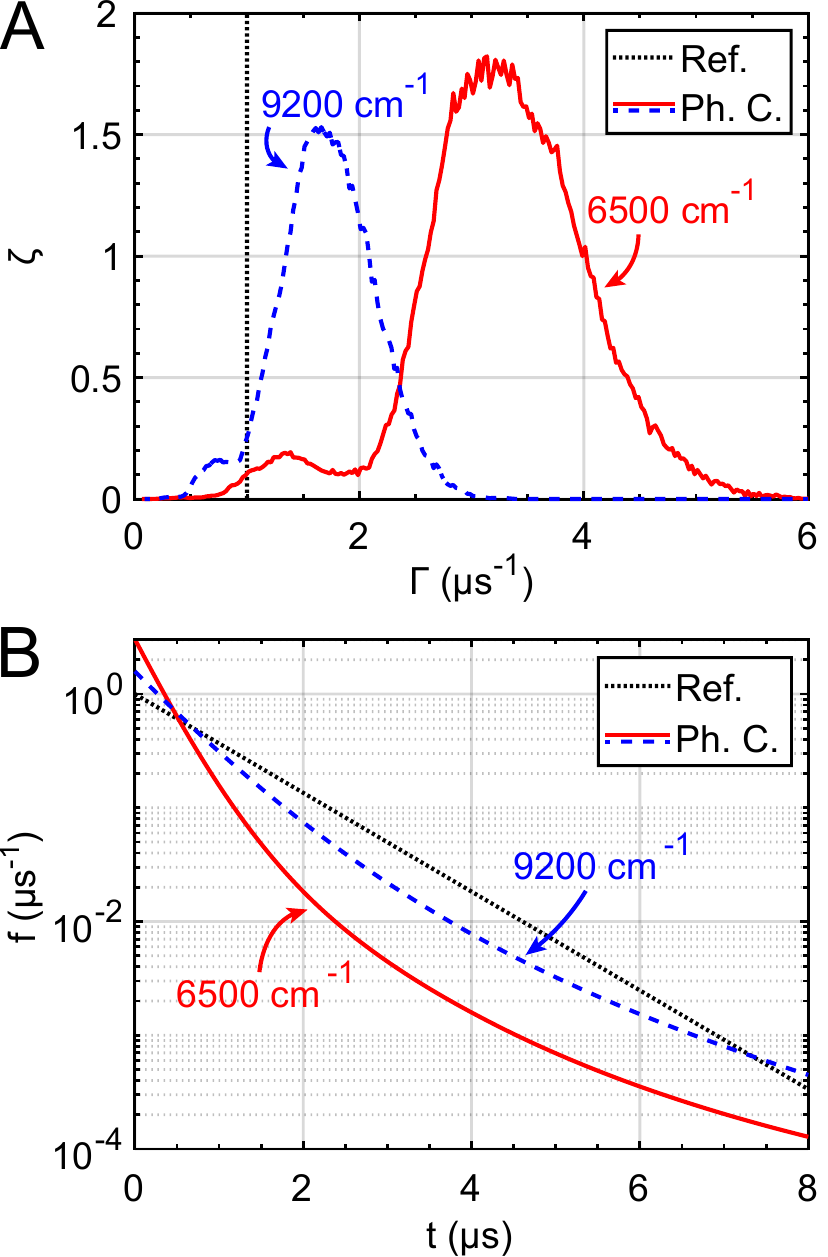}
    \caption{
    (a) The $\zeta(\Gamma)$ distribution calculated at $\tilde{\nu} =$ \pcm{6500\pm100} (red solid) and $\tilde{\nu} =$ \pcm{9200\pm100} (blue dashed) inside the pores of the photonic crystal (`Ph. C.') with an inverse woodpile structure.
    The input for the calculations (red/blue) is the reference $\delta$ distribution (`Ref.', black) with amplitude 1 and decay rate \SI{1}{\per\micro\second}.
    (b) The corresponding decay curves calculated from Eq.~\ref{eq:ldos_ft_to_zetaGamma} on a semi-logarithmic scale.
    The integrals over the total curves are the same.
    }
    \label{fig:ldos_zeta_ft_wnr6500}
\end{figure}

We turn to the computation of the distribution $\zeta(\Gamma)$ and the decay curve $f(t)$ for positions inside the pores of the photonic crystal with an inverse woodpile structure. 
As in many experiments, the calculations of $\zeta(\Gamma)$ and $f(t)$ in the photonic crystal are relative to a decay curve in a homogeneous reference medium (here $n = 1.48$).
We choose a reference decay curve of $f_{\rm{ref}}(t) = A_{\rm{ref}} \Gamma_{\rm{ref}}\exp{-\Gamma_{\rm{ref}} t}$ and corresponding $\zeta(\Gamma) = A_{\rm{ref}}\Gamma_{\rm{ref}}\delta(\Gamma-\Gamma_{\rm{ref}})$ with $A_{\rm{ref}} =$ 1 and $\Gamma_{\rm{ref}} =$ \SI{1}{\per\micro\second}, see the black dotted lines in Fig.~\ref{fig:ldos_zeta_ft_wnr6500}B and~\ref{fig:ldos_zeta_ft_wnr6500}A, respectively.
For simplicity, we assume \pct{100} quantum efficiency, i.e., the non-radiative decay rate is zero, and this can be generalized to any realistic quantum emitter.

We compute $\zeta(\Gamma)$ inside the photonic crystal pores in Fig.~\ref{fig:ldos_zeta_ft_wnr6500}A in two wavenumber intervals: just above the photonic band gap in $\tilde{\nu} =$ \pcm{6500\pm100} (red solid), and further above the band gap in $\tilde{\nu} =$ \pcm{9200\pm100} (blue dashed).
We observe average enhancements in rates $\Gamma$ of 3.3$\!\,\times$ and  1.7$\!\,\times$ relative to the reference, respectively. 
As expected from Fig.~\ref{fig:ldos_avg_iwp_pores}, just above the band gap at $\tilde{\nu} =$ \pcm{6500} the average enhancement is large and the $\zeta(\Gamma)$ distribution is broad, while further above the band gap at \pcm{9200}, the enhancement is lower and the width narrower.
The broader the distribution $\zeta(\Gamma)$ becomes, the more decay rates are present in the ensemble. 

The decay curves $f(t)$ are calculated from $\zeta(\Gamma)$ using Eq.~\ref{eq:ldos_ft_to_zetaGamma} and shown in Fig.~\ref{fig:ldos_zeta_ft_wnr6500}B.
These curves resemble what an experimentalist measures in a time-correlated single-photon counting (TCSPC) experiment probing an ensemble of stationary rotating emitters homogeneously distributed over the pores of an inverse woodpile photonic crystal.
At both 6500 and \pcm{9200}, the resulting slopes differ greatly from linear trends on the semi-logarithmic scale, and therefore, decay is strongly non-single-exponential, as has been experimentally observed~\cite{lodahl2004nature, Nikolaev2007PRB, Vallee2007PRB}.
In the first few microseconds, the slope of the photonic crystal decay traces is steep due to the high decay rates present in $\zeta(\Gamma)$.
In contrast, in the last few microseconds, the reference's slope is steeper than the photonic crystal's because the $\zeta(\Gamma)$ is non-zero at $\Gamma < \Gamma_{\rm{ref}}$.
The slower decay in the last few microseconds is challenging to observe experimentally because the signal is small, which will require a substantial signal-to-noise ratio: $f(t =$ \us{7}$)/f(t = 0)\sim 10^{-4}$.
We note that the time integrals of the curves in Fig.~\ref{fig:ldos_zeta_ft_wnr6500}B are all the same because in time-correlated single-photon counting experiments with emitters that have \pct{100} quantum efficiency, each emitter will eventually radiatively decay.

In experiments, obtaining the distribution $\zeta(\Gamma)$ from $f(t)$ is difficult because it requires an inverse Laplace transform.
A common approach to finding $\zeta(\Gamma)$ is to fit $f(t)$ with a function that preferably has few free parameters, a known inverse Laplace transform, fits many decay curves well, and approaches zero as $t \to \infty$.
We observe that our $\zeta(\Gamma)$ distributions in Fig.~\ref{fig:ldos_zeta_ft_wnr6500}A have a relatively steep edge at low $\Gamma$ and a more gentle edge at high $\Gamma$.
Therefore, the main peak in the $\zeta(\Gamma)$ distribution is reasonably well described by a log-normal distribution, which has only three free parameters, a known Laplace transform, and goes to zero when $t \to \infty$~\cite{vanDriel2007PRB}.
It has already been experimentally observed that the log-normal distribution fits the decay curve of an ensemble of emitters in 3D photonic crystals very well~\cite{Nikolaev2007PRB}; we now also have an understanding of why.



\section{Conclusion}
We have calculated the radiative decay rate in 3D photonic crystals using two different methods: weighted counting of Bloch modes via plane-wave expansion (PWE) and finite-difference time-domain (FDTD) simulations of the power emitted by a dipole.
The methods agree well. 
The slight differences are attributed to the finite-size and surface effects that are present in FDTD, but absent in PWE.
Furthermore, we used the PWE method to calculate the decay curve of an ensemble of emitters distributed inhomogeneously in a 3D photonic band-gap crystal.

Our computations provide new insights into what to expect in experiments.
For example, we have provided an upper bound to the enhancement in decay rate: the enhancement is not much larger than one would obtain using a homogeneous medium of the high refractive index material present in the photonic crystal, except at frequencies just above a 3D band gap.

The results for stationary rotating dipoles can be extended to stationary non-rotating dipoles by also considering the dipole orientation in Eq.~\ref{eq:ldos_ldos}.
Furthermore, the speckle in FDTD simulations can be reduced by using a sphere-shaped structure instead of the box-shaped one we used here.
Lastly, while our results focus on the decay curve, the enhancement and inhibition of the emitted power of an ensemble of emitters in a photonic crystal can also be obtained by considering non-radiative decay.

\begin{acknowledgments}\label{sec:acknowledgments}
We thank Lars Corbijn van Willenswaard for insightful discussions.
This work was supported by the project ``Self-Assembled Icosahedral Photonic Quasicrystals with a Band Gap for Visible Light” (OCENW.GROOT.2019.071) of the ``Nederlandse Organisatie voor Wetenschappelijk Onderzoek'' (NWO); 
the NWO-TTW Perspectief program P15-36 ``Free-form scattering optics'' (FFSO) with TUE, TUD, ASML, Demcon, Lumileds, Schott, Signify, and TNO; 
and by the NWO-TTW Perspectief program P21-20 ``Optical coherence; optimal delivery and positioning'' (OPTIC) with TUE, TUD, ARCNL, Anteryon, ASML, Demcon, JMO, Signify, and TNO.
\end{acknowledgments}

\section*{Data Availability}
The data that support the findings of this article will be openly available upon publishing.

\bibliography{references_RLDOS} 

\begin{thebibliography}{52}%
\makeatletter
\providecommand \@ifxundefined [1]{%
 \@ifx{#1\undefined}
}%
\providecommand \@ifnum [1]{%
 \ifnum #1\expandafter \@firstoftwo
 \else \expandafter \@secondoftwo
 \fi
}%
\providecommand \@ifx [1]{%
 \ifx #1\expandafter \@firstoftwo
 \else \expandafter \@secondoftwo
 \fi
}%
\providecommand \natexlab [1]{#1}%
\providecommand \enquote  [1]{``#1''}%
\providecommand \bibnamefont  [1]{#1}%
\providecommand \bibfnamefont [1]{#1}%
\providecommand \citenamefont [1]{#1}%
\providecommand \href@noop [0]{\@secondoftwo}%
\providecommand \href [0]{\begingroup \@sanitize@url \@href}%
\providecommand \@href[1]{\@@startlink{#1}\@@href}%
\providecommand \@@href[1]{\endgroup#1\@@endlink}%
\providecommand \@sanitize@url [0]{\catcode `\\12\catcode `\$12\catcode `\&12\catcode `\#12\catcode `\^12\catcode `\_12\catcode `\%12\relax}%
\providecommand \@@startlink[1]{}%
\providecommand \@@endlink[0]{}%
\providecommand \url  [0]{\begingroup\@sanitize@url \@url }%
\providecommand \@url [1]{\endgroup\@href {#1}{\urlprefix }}%
\providecommand \urlprefix  [0]{URL }%
\providecommand \Eprint [0]{\href }%
\providecommand \doibase [0]{https://doi.org/}%
\providecommand \selectlanguage [0]{\@gobble}%
\providecommand \bibinfo  [0]{\@secondoftwo}%
\providecommand \bibfield  [0]{\@secondoftwo}%
\providecommand \translation [1]{[#1]}%
\providecommand \BibitemOpen [0]{}%
\providecommand \bibitemStop [0]{}%
\providecommand \bibitemNoStop [0]{.\EOS\space}%
\providecommand \EOS [0]{\spacefactor3000\relax}%
\providecommand \BibitemShut  [1]{\csname bibitem#1\endcsname}%
\let\auto@bib@innerbib\@empty
\bibitem [{\citenamefont {Novotny}\ and\ \citenamefont {Hecht}(2012)}]{Novotny2012book}%
  \BibitemOpen
  \bibfield  {author} {\bibinfo {author} {\bibfnamefont {L.}~\bibnamefont {Novotny}}\ and\ \bibinfo {author} {\bibfnamefont {B.}~\bibnamefont {Hecht}},\ }\href@noop {} {\emph {\bibinfo {title} {{Principles of Nano-Optics}}}},\ \bibinfo {edition} {2nd}\ ed.\ (\bibinfo  {publisher} {Cambridge University},\ \bibinfo {address} {Cambridge},\ \bibinfo {year} {2012})\BibitemShut {NoStop}%
\bibitem [{\citenamefont {Hewitt}(2015)}]{hewitt2015book}%
  \BibitemOpen
  \bibfield  {author} {\bibinfo {author} {\bibfnamefont {P.~G.}\ \bibnamefont {Hewitt}},\ }\href@noop {} {\emph {\bibinfo {title} {{Conceptual Physics}}}}\ (\bibinfo  {publisher} {Pearson},\ \bibinfo {address} {Edinburgh},\ \bibinfo {year} {2015})\BibitemShut {NoStop}%
\bibitem [{\citenamefont {Milonni}(1994)}]{Milonni1994bookQV}%
  \BibitemOpen
  \bibfield  {author} {\bibinfo {author} {\bibfnamefont {P.~W.}\ \bibnamefont {Milonni}},\ }\href@noop {} {\emph {\bibinfo {title} {{The Quantum Vacuum: an Introduction to Quantum Electrodynamics}}}}\ (\bibinfo  {publisher} {Academic Press},\ \bibinfo {address} {Boston},\ \bibinfo {year} {1994})\BibitemShut {NoStop}%
\bibitem [{\citenamefont {Bykov}(1972)}]{Bykov1972JETP}%
  \BibitemOpen
  \bibfield  {author} {\bibinfo {author} {\bibfnamefont {V.~P.}\ \bibnamefont {Bykov}},\ }\bibinfo {title} {{Spontaneous Emission in a Periodic Structure}},\ \href@noop {} {\bibfield  {journal} {\bibinfo  {journal} {Sov. Phys. JETP}\ }\textbf {\bibinfo {volume} {35}},\ \bibinfo {pages} {269} (\bibinfo {year} {1972})}\BibitemShut {NoStop}%
\bibitem [{\citenamefont {Yablonovitch}(1987)}]{Yablonovitch1987PRL}%
  \BibitemOpen
  \bibfield  {author} {\bibinfo {author} {\bibfnamefont {E.}~\bibnamefont {Yablonovitch}},\ }\bibinfo {title} {{Inhibited Spontaneous Emission in Solid-State Physics and Electronics}},\ \href {https://doi.org/10.1103/PhysRevLett.58.2059} {\bibfield  {journal} {\bibinfo  {journal} {Phys. Rev. Lett.}\ }\textbf {\bibinfo {volume} {58}},\ \bibinfo {pages} {2059} (\bibinfo {year} {1987})}\BibitemShut {NoStop}%
\bibitem [{\citenamefont {Noda}\ \emph {et~al.}(2000)\citenamefont {Noda}, \citenamefont {Tomoda}, \citenamefont {Yamamoto},\ and\ \citenamefont {Chutinan}}]{noda2000science}%
  \BibitemOpen
  \bibfield  {author} {\bibinfo {author} {\bibfnamefont {S.}~\bibnamefont {Noda}}, \bibinfo {author} {\bibfnamefont {K.}~\bibnamefont {Tomoda}}, \bibinfo {author} {\bibfnamefont {N.}~\bibnamefont {Yamamoto}},\ and\ \bibinfo {author} {\bibfnamefont {A.}~\bibnamefont {Chutinan}},\ }\bibinfo {title} {{Full Three-Dimensional Photonic Bandgap Crystals at Near-Infrared Wavelengths}},\ \href@noop {} {\bibfield  {journal} {\bibinfo  {journal} {Science}\ }\textbf {\bibinfo {volume} {289}},\ \bibinfo {pages} {604} (\bibinfo {year} {2000})}\BibitemShut {NoStop}%
\bibitem [{\citenamefont {Vats}\ \emph {et~al.}(2002)\citenamefont {Vats}, \citenamefont {John},\ and\ \citenamefont {Busch}}]{Vats2002PRA}%
  \BibitemOpen
  \bibfield  {author} {\bibinfo {author} {\bibfnamefont {N.}~\bibnamefont {Vats}}, \bibinfo {author} {\bibfnamefont {S.}~\bibnamefont {John}},\ and\ \bibinfo {author} {\bibfnamefont {K.}~\bibnamefont {Busch}},\ }\bibinfo {title} {{Theory of fluorescence in photonic crystals}},\ \href {https://doi.org/10.1103/PhysRevA.65.043808} {\bibfield  {journal} {\bibinfo  {journal} {Phys. Rev. A}\ }\textbf {\bibinfo {volume} {65}},\ \bibinfo {pages} {043808} (\bibinfo {year} {2002})}\BibitemShut {NoStop}%
\bibitem [{\citenamefont {Leistikow}\ \emph {et~al.}(2011)\citenamefont {Leistikow}, \citenamefont {Mosk}, \citenamefont {Yeganegi}, \citenamefont {Huisman}, \citenamefont {Lagendijk},\ and\ \citenamefont {Vos}}]{Leistikow2011PRL}%
  \BibitemOpen
  \bibfield  {author} {\bibinfo {author} {\bibfnamefont {M.~D.}\ \bibnamefont {Leistikow}}, \bibinfo {author} {\bibfnamefont {A.~P.}\ \bibnamefont {Mosk}}, \bibinfo {author} {\bibfnamefont {E.}~\bibnamefont {Yeganegi}}, \bibinfo {author} {\bibfnamefont {S.~R.}\ \bibnamefont {Huisman}}, \bibinfo {author} {\bibfnamefont {A.}~\bibnamefont {Lagendijk}},\ and\ \bibinfo {author} {\bibfnamefont {W.~L.}\ \bibnamefont {Vos}},\ }\bibinfo {title} {{Inhibited Spontaneous Emission of Quantum Dots Observed in a 3D Photonic Band Gap}},\ \href@noop {} {\bibfield  {journal} {\bibinfo  {journal} {Phys. Rev. Lett.}\ }\textbf {\bibinfo {volume} {107}},\ \bibinfo {pages} {193903} (\bibinfo {year} {2011})}\BibitemShut {NoStop}%
\bibitem [{\citenamefont {Joannopoulos}\ \emph {et~al.}(2008)\citenamefont {Joannopoulos}, \citenamefont {Johnson}, \citenamefont {Winn},\ and\ \citenamefont {Meade}}]{Joannopoulos2008book}%
  \BibitemOpen
  \bibfield  {author} {\bibinfo {author} {\bibfnamefont {J.}~\bibnamefont {Joannopoulos}}, \bibinfo {author} {\bibfnamefont {S.~G.}\ \bibnamefont {Johnson}}, \bibinfo {author} {\bibfnamefont {J.}~\bibnamefont {Winn}},\ and\ \bibinfo {author} {\bibfnamefont {R.~D.}\ \bibnamefont {Meade}},\ }\href@noop {} {\emph {\bibinfo {title} {{Photonic Crystals: Molding the Flow of Light}}}},\ \bibinfo {edition} {2nd}\ ed.\ (\bibinfo  {publisher} {Princeton University Press},\ \bibinfo {address} {Princeton},\ \bibinfo {year} {2008})\BibitemShut {NoStop}%
\bibitem [{\citenamefont {Fermi}(1932)}]{Fermi1932RMP}%
  \BibitemOpen
  \bibfield  {author} {\bibinfo {author} {\bibfnamefont {E.}~\bibnamefont {Fermi}},\ }\bibinfo {title} {{Quantum Theory of Radiation}},\ \href {https://doi.org/10.1103/RevModPhys.4.87} {\bibfield  {journal} {\bibinfo  {journal} {Rev. Mod. Phys.}\ }\textbf {\bibinfo {volume} {4}},\ \bibinfo {pages} {87} (\bibinfo {year} {1932})}\BibitemShut {NoStop}%
\bibitem [{\citenamefont {Snoeks}\ \emph {et~al.}(1995)\citenamefont {Snoeks}, \citenamefont {Lagendijk},\ and\ \citenamefont {Polman}}]{Snoeks1995PRL}%
  \BibitemOpen
  \bibfield  {author} {\bibinfo {author} {\bibfnamefont {E.}~\bibnamefont {Snoeks}}, \bibinfo {author} {\bibfnamefont {A.}~\bibnamefont {Lagendijk}},\ and\ \bibinfo {author} {\bibfnamefont {A.}~\bibnamefont {Polman}},\ }\bibinfo {title} {{Measuring and Modifying the Spontaneous Emission Rate of Erbium near an Interface}},\ \href {https://doi.org/10.1103/PhysRevLett.74.2459} {\bibfield  {journal} {\bibinfo  {journal} {Phys. Rev. Lett.}\ }\textbf {\bibinfo {volume} {74}},\ \bibinfo {pages} {2459} (\bibinfo {year} {1995})}\BibitemShut {NoStop}%
\bibitem [{\citenamefont {Asatryan}\ \emph {et~al.}(2001)\citenamefont {Asatryan}, \citenamefont {Busch}, \citenamefont {McPhedran}, \citenamefont {Botten}, \citenamefont {de~Sterke},\ and\ \citenamefont {Nicorovici}}]{asatryan2001PRE}%
  \BibitemOpen
  \bibfield  {author} {\bibinfo {author} {\bibfnamefont {A.~A.}\ \bibnamefont {Asatryan}}, \bibinfo {author} {\bibfnamefont {K.}~\bibnamefont {Busch}}, \bibinfo {author} {\bibfnamefont {R.~C.}\ \bibnamefont {McPhedran}}, \bibinfo {author} {\bibfnamefont {L.~C.}\ \bibnamefont {Botten}}, \bibinfo {author} {\bibfnamefont {C.~M.}\ \bibnamefont {de~Sterke}},\ and\ \bibinfo {author} {\bibfnamefont {N.~A.}\ \bibnamefont {Nicorovici}},\ }\bibinfo {title} {Two-dimensional green’s function and local density of states in photonic crystals consisting of a finite number of cylinders of infinite length},\ \href@noop {} {\bibfield  {journal} {\bibinfo  {journal} {Phys. Rev. E}\ }\textbf {\bibinfo {volume} {63}},\ \bibinfo {pages} {046612} (\bibinfo {year} {2001})}\BibitemShut {NoStop}%
\bibitem [{\citenamefont {Rockstuhl}\ and\ \citenamefont {Lederer}(2009)}]{rockstuhl2009PRB}%
  \BibitemOpen
  \bibfield  {author} {\bibinfo {author} {\bibfnamefont {C.}~\bibnamefont {Rockstuhl}}\ and\ \bibinfo {author} {\bibfnamefont {F.}~\bibnamefont {Lederer}},\ }\bibinfo {title} {Suppression of the local density of states in a medium made of randomly arranged dielectric spheres},\ \href@noop {} {\bibfield  {journal} {\bibinfo  {journal} {Phys. Rev. B}\ }\textbf {\bibinfo {volume} {79}},\ \bibinfo {pages} {132202} (\bibinfo {year} {2009})}\BibitemShut {NoStop}%
\bibitem [{\citenamefont {Vos}\ \emph {et~al.}(2009)\citenamefont {Vos}, \citenamefont {Koenderink},\ and\ \citenamefont {Nikolaev}}]{vos2009PRA}%
  \BibitemOpen
  \bibfield  {author} {\bibinfo {author} {\bibfnamefont {W.~L.}\ \bibnamefont {Vos}}, \bibinfo {author} {\bibfnamefont {A.~F.}\ \bibnamefont {Koenderink}},\ and\ \bibinfo {author} {\bibfnamefont {I.~S.}\ \bibnamefont {Nikolaev}},\ }\bibinfo {title} {{Orientation-dependent spontaneous emission rates of a two-level quantum emitter in any nanophotonic environment}},\ \href@noop {} {\bibfield  {journal} {\bibinfo  {journal} {Phys. Rev. A}\ }\textbf {\bibinfo {volume} {80}},\ \bibinfo {pages} {053802} (\bibinfo {year} {2009})}\BibitemShut {NoStop}%
\bibitem [{\citenamefont {Lodahl}\ \emph {et~al.}(2015)\citenamefont {Lodahl}, \citenamefont {Mahmoodian},\ and\ \citenamefont {Stobbe}}]{Lodahl2015RevModPhys}%
  \BibitemOpen
  \bibfield  {author} {\bibinfo {author} {\bibfnamefont {P.}~\bibnamefont {Lodahl}}, \bibinfo {author} {\bibfnamefont {S.}~\bibnamefont {Mahmoodian}},\ and\ \bibinfo {author} {\bibfnamefont {S.}~\bibnamefont {Stobbe}},\ }\bibinfo {title} {{Interfacing single photons and single quantum dots with photonic nanostructures}},\ \href {https://doi.org/10.1103/RevModPhys.87.347} {\bibfield  {journal} {\bibinfo  {journal} {Rev. Mod. Phys.}\ }\textbf {\bibinfo {volume} {87}},\ \bibinfo {pages} {347} (\bibinfo {year} {2015})}\BibitemShut {NoStop}%
\bibitem [{\citenamefont {Nikolaev}\ \emph {et~al.}(2009)\citenamefont {Nikolaev}, \citenamefont {Vos},\ and\ \citenamefont {Koenderink}}]{Nikolaev2009JOSAB}%
  \BibitemOpen
  \bibfield  {author} {\bibinfo {author} {\bibfnamefont {I.~S.}\ \bibnamefont {Nikolaev}}, \bibinfo {author} {\bibfnamefont {W.~L.}\ \bibnamefont {Vos}},\ and\ \bibinfo {author} {\bibfnamefont {A.~F.}\ \bibnamefont {Koenderink}},\ }\bibinfo {title} {{Accurate calculation of the local density of optical states in inverse-opal photonic crystals}},\ \href {https://doi.org/10.1364/JOSAB.26.000987} {\bibfield  {journal} {\bibinfo  {journal} {J. Opt. Soc. Am. B}\ }\textbf {\bibinfo {volume} {26}},\ \bibinfo {pages} {987} (\bibinfo {year} {2009})}\BibitemShut {NoStop}%
\bibitem [{1()}]{1}%
  \BibitemOpen
  \href@noop {} {}\bibinfo {note} {To use the RLDOS, the dipole should be rotating fast enough to average out directional variations in the---later specified---partial RLDOS.}\BibitemShut {Stop}%
\bibitem [{\citenamefont {Sprik}\ \emph {et~al.}(1996)\citenamefont {Sprik}, \citenamefont {van Tiggelen},\ and\ \citenamefont {Lagendijk}}]{Sprik1996EPL}%
  \BibitemOpen
  \bibfield  {author} {\bibinfo {author} {\bibfnamefont {R.}~\bibnamefont {Sprik}}, \bibinfo {author} {\bibfnamefont {B.~A.}\ \bibnamefont {van Tiggelen}},\ and\ \bibinfo {author} {\bibfnamefont {A.}~\bibnamefont {Lagendijk}},\ }\bibinfo {title} {{Optical emission in periodic dielectrics}},\ \href {http://stacks.iop.org/0295-5075/35/i=4/a=265} {\bibfield  {journal} {\bibinfo  {journal} {Europhys. Lett.}\ }\textbf {\bibinfo {volume} {35}},\ \bibinfo {pages} {265} (\bibinfo {year} {1996})}\BibitemShut {NoStop}%
\bibitem [{\citenamefont {Busch}\ and\ \citenamefont {John}(1998)}]{Busch1998PRE}%
  \BibitemOpen
  \bibfield  {author} {\bibinfo {author} {\bibfnamefont {K.}~\bibnamefont {Busch}}\ and\ \bibinfo {author} {\bibfnamefont {S.}~\bibnamefont {John}},\ }\bibinfo {title} {{Photonic band gap formation in certain self-organizing systems}},\ \href {https://doi.org/10.1103/PhysRevE.58.3896} {\bibfield  {journal} {\bibinfo  {journal} {Phys. Rev. E}\ }\textbf {\bibinfo {volume} {58}},\ \bibinfo {pages} {3896} (\bibinfo {year} {1998})}\BibitemShut {NoStop}%
\bibitem [{\citenamefont {Li}\ \emph {et~al.}(2000)\citenamefont {Li}, \citenamefont {Lin},\ and\ \citenamefont {Zhang}}]{li2000PRL}%
  \BibitemOpen
  \bibfield  {author} {\bibinfo {author} {\bibfnamefont {Z.-Y.}\ \bibnamefont {Li}}, \bibinfo {author} {\bibfnamefont {L.-L.}\ \bibnamefont {Lin}},\ and\ \bibinfo {author} {\bibfnamefont {Z.-Q.}\ \bibnamefont {Zhang}},\ }\bibinfo {title} {{Spontaneous Emission from Photonic Crystals: Full Vectorial Calculations}},\ \href@noop {} {\bibfield  {journal} {\bibinfo  {journal} {Phys. Rev. Lett.}\ }\textbf {\bibinfo {volume} {84}},\ \bibinfo {pages} {4341} (\bibinfo {year} {2000})}\BibitemShut {NoStop}%
\bibitem [{\citenamefont {Wang}\ \emph {et~al.}(2002)\citenamefont {Wang}, \citenamefont {Wang}, \citenamefont {Gu},\ and\ \citenamefont {Yang}}]{wang2002PRL}%
  \BibitemOpen
  \bibfield  {author} {\bibinfo {author} {\bibfnamefont {X.-H.}\ \bibnamefont {Wang}}, \bibinfo {author} {\bibfnamefont {R.}~\bibnamefont {Wang}}, \bibinfo {author} {\bibfnamefont {B.-Y.}\ \bibnamefont {Gu}},\ and\ \bibinfo {author} {\bibfnamefont {G.-Z.}\ \bibnamefont {Yang}},\ }\bibinfo {title} {{Decay Distribution of Spontaneous Emission from an Assembly of Atoms in Photonic Crystals with Pseudogaps}},\ \href@noop {} {\bibfield  {journal} {\bibinfo  {journal} {Phys. Rev. Lett.}\ }\textbf {\bibinfo {volume} {88}},\ \bibinfo {pages} {093902} (\bibinfo {year} {2002})}\BibitemShut {NoStop}%
\bibitem [{\citenamefont {Wang}\ \emph {et~al.}(2003)\citenamefont {Wang}, \citenamefont {Wang}, \citenamefont {Gu},\ and\ \citenamefont {Yang}}]{wang2003PRB}%
  \BibitemOpen
  \bibfield  {author} {\bibinfo {author} {\bibfnamefont {R.}~\bibnamefont {Wang}}, \bibinfo {author} {\bibfnamefont {X.-H.}\ \bibnamefont {Wang}}, \bibinfo {author} {\bibfnamefont {B.-Y.}\ \bibnamefont {Gu}},\ and\ \bibinfo {author} {\bibfnamefont {G.-Z.}\ \bibnamefont {Yang}},\ }\bibinfo {title} {{Local density of states in three-dimensional photonic crystals: calculation and enhancement effects}},\ \href@noop {} {\bibfield  {journal} {\bibinfo  {journal} {Phys. Rev. B}\ }\textbf {\bibinfo {volume} {67}},\ \bibinfo {pages} {155114} (\bibinfo {year} {2003})}\BibitemShut {NoStop}%
\bibitem [{\citenamefont {Taflove}\ \emph {et~al.}(2013)\citenamefont {Taflove}, \citenamefont {Oskooi},\ and\ \citenamefont {Johnson}}]{taflove2005book}%
  \BibitemOpen
  \bibfield  {author} {\bibinfo {author} {\bibfnamefont {A.}~\bibnamefont {Taflove}}, \bibinfo {author} {\bibfnamefont {A.}~\bibnamefont {Oskooi}},\ and\ \bibinfo {author} {\bibfnamefont {S.}~\bibnamefont {Johnson}},\ }\href@noop {} {\emph {\bibinfo {title} {{Advances in FDTD Computational Electrodynamics: Photonics and Nanotechnology}}}}\ (\bibinfo  {publisher} {Elsevier},\ \bibinfo {address} {Amsterdam},\ \bibinfo {year} {2013})\BibitemShut {NoStop}%
\bibitem [{\citenamefont {Hermann}\ and\ \citenamefont {Hess}(2002)}]{hermann2002JOSAB}%
  \BibitemOpen
  \bibfield  {author} {\bibinfo {author} {\bibfnamefont {C.}~\bibnamefont {Hermann}}\ and\ \bibinfo {author} {\bibfnamefont {O.}~\bibnamefont {Hess}},\ }\bibinfo {title} {{Modified spontaneous-emission rate in an inverted-opal structure with complete photonic bandgap}},\ \href@noop {} {\bibfield  {journal} {\bibinfo  {journal} {J. Opt. Soc. Am. B}\ }\textbf {\bibinfo {volume} {19}},\ \bibinfo {pages} {3013} (\bibinfo {year} {2002})}\BibitemShut {NoStop}%
\bibitem [{\citenamefont {Ishizaki}\ \emph {et~al.}(2009)\citenamefont {Ishizaki}, \citenamefont {Okano},\ and\ \citenamefont {Noda}}]{ishizaki2009JOSAB}%
  \BibitemOpen
  \bibfield  {author} {\bibinfo {author} {\bibfnamefont {K.}~\bibnamefont {Ishizaki}}, \bibinfo {author} {\bibfnamefont {M.}~\bibnamefont {Okano}},\ and\ \bibinfo {author} {\bibfnamefont {S.}~\bibnamefont {Noda}},\ }\bibinfo {title} {{Numerical investigation of emission in finite-sized, three-dimensional photonic crystals with structural fluctuations}},\ \href@noop {} {\bibfield  {journal} {\bibinfo  {journal} {J. Opt. Soc. Am. B}\ }\textbf {\bibinfo {volume} {26}},\ \bibinfo {pages} {1157} (\bibinfo {year} {2009})}\BibitemShut {NoStop}%
\bibitem [{\citenamefont {Mavidis}\ \emph {et~al.}(2020)\citenamefont {Mavidis}, \citenamefont {Tasolamprou}, \citenamefont {Hasan}, \citenamefont {Koschny}, \citenamefont {Economou}, \citenamefont {Kafesaki}, \citenamefont {Soukoulis},\ and\ \citenamefont {Vos}}]{Mavidis2020PRB}%
  \BibitemOpen
  \bibfield  {author} {\bibinfo {author} {\bibfnamefont {C.~P.}\ \bibnamefont {Mavidis}}, \bibinfo {author} {\bibfnamefont {A.~C.}\ \bibnamefont {Tasolamprou}}, \bibinfo {author} {\bibfnamefont {S.~B.}\ \bibnamefont {Hasan}}, \bibinfo {author} {\bibfnamefont {T.}~\bibnamefont {Koschny}}, \bibinfo {author} {\bibfnamefont {E.~N.}\ \bibnamefont {Economou}}, \bibinfo {author} {\bibfnamefont {M.}~\bibnamefont {Kafesaki}}, \bibinfo {author} {\bibfnamefont {C.~M.}\ \bibnamefont {Soukoulis}},\ and\ \bibinfo {author} {\bibfnamefont {W.~L.}\ \bibnamefont {Vos}},\ }\bibinfo {title} {{The local density of optical states in the 3D band gap of a finite photonic crystal}},\ \href {https://doi.org/10.1103/PhysRevB.101.235309} {\bibfield  {journal} {\bibinfo  {journal} {Phys. Rev. B}\ }\textbf {\bibinfo {volume} {101}},\ \bibinfo {pages} {235309} (\bibinfo {year} {2020})}\BibitemShut {NoStop}%
\bibitem [{\citenamefont {Xu}\ \emph {et~al.}(2000)\citenamefont {Xu}, \citenamefont {Lee},\ and\ \citenamefont {Yariv}}]{xu2000pra}%
  \BibitemOpen
  \bibfield  {author} {\bibinfo {author} {\bibfnamefont {Y.}~\bibnamefont {Xu}}, \bibinfo {author} {\bibfnamefont {R.~K.}\ \bibnamefont {Lee}},\ and\ \bibinfo {author} {\bibfnamefont {A.}~\bibnamefont {Yariv}},\ }\bibinfo {title} {{Quantum analysis and the classical analysis of spontaneous emission in a microcavity}},\ \href@noop {} {\bibfield  {journal} {\bibinfo  {journal} {Phys. Rev. A}\ }\textbf {\bibinfo {volume} {61}},\ \bibinfo {pages} {033807} (\bibinfo {year} {2000})}\BibitemShut {NoStop}%
\bibitem [{\citenamefont {Petrov}\ \emph {et~al.}(1998)\citenamefont {Petrov}, \citenamefont {Bogomolov}, \citenamefont {Kalosha},\ and\ \citenamefont {Gaponenko}}]{petrov1998PRL}%
  \BibitemOpen
  \bibfield  {author} {\bibinfo {author} {\bibfnamefont {E.~P.}\ \bibnamefont {Petrov}}, \bibinfo {author} {\bibfnamefont {V.~N.}\ \bibnamefont {Bogomolov}}, \bibinfo {author} {\bibfnamefont {I.~I.}\ \bibnamefont {Kalosha}},\ and\ \bibinfo {author} {\bibfnamefont {S.~V.}\ \bibnamefont {Gaponenko}},\ }\bibinfo {title} {{Spontaneous emission of organic molecules embedded in a photonic crystal}},\ \href@noop {} {\bibfield  {journal} {\bibinfo  {journal} {Phys. Rev. Lett.}\ }\textbf {\bibinfo {volume} {81}},\ \bibinfo {pages} {77} (\bibinfo {year} {1998})}\BibitemShut {NoStop}%
\bibitem [{\citenamefont {Yoshino}\ \emph {et~al.}(1998)\citenamefont {Yoshino}, \citenamefont {Lee}, \citenamefont {Tatsuhara}, \citenamefont {Kawagishi}, \citenamefont {Ozaki},\ and\ \citenamefont {Zakhidov}}]{yoshino1998APL}%
  \BibitemOpen
  \bibfield  {author} {\bibinfo {author} {\bibfnamefont {K.}~\bibnamefont {Yoshino}}, \bibinfo {author} {\bibfnamefont {S.~B.}\ \bibnamefont {Lee}}, \bibinfo {author} {\bibfnamefont {S.}~\bibnamefont {Tatsuhara}}, \bibinfo {author} {\bibfnamefont {Y.}~\bibnamefont {Kawagishi}}, \bibinfo {author} {\bibfnamefont {M.~Z. A.~A.}\ \bibnamefont {Ozaki}},\ and\ \bibinfo {author} {\bibfnamefont {A.~A.}\ \bibnamefont {Zakhidov}},\ }\bibinfo {title} {{Observation of inhibited spontaneous emission and stimulated emission of rhodamine 6G in polymer replica of synthetic opal}},\ \href@noop {} {\bibfield  {journal} {\bibinfo  {journal} {Appl. Phys. Lett.}\ }\textbf {\bibinfo {volume} {73}},\ \bibinfo {pages} {3506} (\bibinfo {year} {1998})}\BibitemShut {NoStop}%
\bibitem [{\citenamefont {Megens}\ \emph {et~al.}(1999)\citenamefont {Megens}, \citenamefont {Schriemer}, \citenamefont {Lagendijk},\ and\ \citenamefont {Vos}}]{megens1999commentPetrov}%
  \BibitemOpen
  \bibfield  {author} {\bibinfo {author} {\bibfnamefont {M.}~\bibnamefont {Megens}}, \bibinfo {author} {\bibfnamefont {H.~P.}\ \bibnamefont {Schriemer}}, \bibinfo {author} {\bibfnamefont {A.}~\bibnamefont {Lagendijk}},\ and\ \bibinfo {author} {\bibfnamefont {W.~L.}\ \bibnamefont {Vos}},\ }\bibinfo {title} {{Comment on ``Spontaneous Emission of Organic Molecules Embedded in a Photonic Crystal''}},\ \href@noop {} {\bibfield  {journal} {\bibinfo  {journal} {Phys. Rev. Lett.}\ }\textbf {\bibinfo {volume} {83}},\ \bibinfo {pages} {5401} (\bibinfo {year} {1999})}\BibitemShut {NoStop}%
\bibitem [{\citenamefont {Lodahl}\ \emph {et~al.}(2004)\citenamefont {Lodahl}, \citenamefont {van Driel}, \citenamefont {Nikolaev}, \citenamefont {Irman}, \citenamefont {Overgaag}, \citenamefont {Vanmaekelbergh},\ and\ \citenamefont {Vos}}]{lodahl2004nature}%
  \BibitemOpen
  \bibfield  {author} {\bibinfo {author} {\bibfnamefont {P.}~\bibnamefont {Lodahl}}, \bibinfo {author} {\bibfnamefont {A.~F.}\ \bibnamefont {van Driel}}, \bibinfo {author} {\bibfnamefont {I.~S.}\ \bibnamefont {Nikolaev}}, \bibinfo {author} {\bibfnamefont {A.}~\bibnamefont {Irman}}, \bibinfo {author} {\bibfnamefont {K.}~\bibnamefont {Overgaag}}, \bibinfo {author} {\bibfnamefont {D.}~\bibnamefont {Vanmaekelbergh}},\ and\ \bibinfo {author} {\bibfnamefont {W.~L.}\ \bibnamefont {Vos}},\ }\bibinfo {title} {{Controlling the dynamics of spontaneous emission from quantum dots by photonic crystals}},\ \href@noop {} {\bibfield  {journal} {\bibinfo  {journal} {Nature (London)}\ }\textbf {\bibinfo {volume} {430}},\ \bibinfo {pages} {654} (\bibinfo {year} {2004})}\BibitemShut {NoStop}%
\bibitem [{\citenamefont {Li}\ \emph {et~al.}(2007)\citenamefont {Li}, \citenamefont {Jia}, \citenamefont {Zhou}, \citenamefont {Bullen}, \citenamefont {Serbin},\ and\ \citenamefont {Gu}}]{Li2007AdvMat}%
  \BibitemOpen
  \bibfield  {author} {\bibinfo {author} {\bibfnamefont {J.}~\bibnamefont {Li}}, \bibinfo {author} {\bibfnamefont {B.}~\bibnamefont {Jia}}, \bibinfo {author} {\bibfnamefont {G.}~\bibnamefont {Zhou}}, \bibinfo {author} {\bibfnamefont {C.}~\bibnamefont {Bullen}}, \bibinfo {author} {\bibfnamefont {J.}~\bibnamefont {Serbin}},\ and\ \bibinfo {author} {\bibfnamefont {M.}~\bibnamefont {Gu}},\ }\bibinfo {title} {{Spectral Redistribution in Spontaneous Emission from Quantum-Dot-Infiltrated 3D Woodpile Photonic Crystals for Telecommunications}},\ \href {https://doi.org/https://doi.org/10.1002/adma.200602054} {\bibfield  {journal} {\bibinfo  {journal} {Adv. Mater.}\ }\textbf {\bibinfo {volume} {19}},\ \bibinfo {pages} {3276} (\bibinfo {year} {2007})}\BibitemShut {NoStop}%
\bibitem [{\citenamefont {Ventura}\ and\ \citenamefont {Gu}(2008)}]{ventura2008AdvMater}%
  \BibitemOpen
  \bibfield  {author} {\bibinfo {author} {\bibfnamefont {M.~J.}\ \bibnamefont {Ventura}}\ and\ \bibinfo {author} {\bibfnamefont {M.}~\bibnamefont {Gu}},\ }\bibinfo {title} {{Engineering Spontaneous Emission in a Quantum-Dot-Doped Polymer Nanocomposite with Three-Dimensional Photonic Crystals}},\ \href@noop {} {\bibfield  {journal} {\bibinfo  {journal} {Adv. Mater.}\ }\textbf {\bibinfo {volume} {20}},\ \bibinfo {pages} {1329} (\bibinfo {year} {2008})}\BibitemShut {NoStop}%
\bibitem [{\citenamefont {Nikolaev}\ \emph {et~al.}(2007)\citenamefont {Nikolaev}, \citenamefont {Lodahl}, \citenamefont {van Driel}, \citenamefont {Koenderink},\ and\ \citenamefont {Vos}}]{Nikolaev2007PRB}%
  \BibitemOpen
  \bibfield  {author} {\bibinfo {author} {\bibfnamefont {I.~S.}\ \bibnamefont {Nikolaev}}, \bibinfo {author} {\bibfnamefont {P.}~\bibnamefont {Lodahl}}, \bibinfo {author} {\bibfnamefont {A.~F.}\ \bibnamefont {van Driel}}, \bibinfo {author} {\bibfnamefont {A.~F.}\ \bibnamefont {Koenderink}},\ and\ \bibinfo {author} {\bibfnamefont {W.~L.}\ \bibnamefont {Vos}},\ }\bibinfo {title} {{Strongly nonexponential time-resolved fluorescence of quantum-dot ensembles in three-dimensional photonic crystals}},\ \href@noop {} {\bibfield  {journal} {\bibinfo  {journal} {Phys. Rev. B}\ }\textbf {\bibinfo {volume} {75}},\ \bibinfo {pages} {115302} (\bibinfo {year} {2007})}\BibitemShut {NoStop}%
\bibitem [{\citenamefont {Vall{\'{e}}e}\ \emph {et~al.}(2007)\citenamefont {Vall{\'{e}}e}, \citenamefont {Baert}, \citenamefont {Kolaric}, \citenamefont {{Van der Auweraer}},\ and\ \citenamefont {Clays}}]{Vallee2007PRB}%
  \BibitemOpen
  \bibfield  {author} {\bibinfo {author} {\bibfnamefont {R.~A.~L.}\ \bibnamefont {Vall{\'{e}}e}}, \bibinfo {author} {\bibfnamefont {K.}~\bibnamefont {Baert}}, \bibinfo {author} {\bibfnamefont {B.}~\bibnamefont {Kolaric}}, \bibinfo {author} {\bibfnamefont {M.}~\bibnamefont {{Van der Auweraer}}},\ and\ \bibinfo {author} {\bibfnamefont {K.}~\bibnamefont {Clays}},\ }\bibinfo {title} {{Nonexponential decay of spontaneous emission from an ensemble of molecules in photonic crystals}},\ \href {https://doi.org/10.1103/PhysRevB.76.045113} {\bibfield  {journal} {\bibinfo  {journal} {Phys. Rev. B}\ }\textbf {\bibinfo {volume} {76}},\ \bibinfo {pages} {045113} (\bibinfo {year} {2007})}\BibitemShut {NoStop}%
\bibitem [{\citenamefont {Vion}\ \emph {et~al.}(2009)\citenamefont {Vion}, \citenamefont {Barthou}, \citenamefont {B{\'{e}}nalloul}, \citenamefont {Schwob}, \citenamefont {Coolen}, \citenamefont {Gruzintev}, \citenamefont {Emel'chenko}, \citenamefont {Masalov}, \citenamefont {Frigerio},\ and\ \citenamefont {Ma{\^{i}}tre}}]{Vion2009JAP}%
  \BibitemOpen
  \bibfield  {author} {\bibinfo {author} {\bibfnamefont {C.}~\bibnamefont {Vion}}, \bibinfo {author} {\bibfnamefont {C.}~\bibnamefont {Barthou}}, \bibinfo {author} {\bibfnamefont {P.}~\bibnamefont {B{\'{e}}nalloul}}, \bibinfo {author} {\bibfnamefont {C.}~\bibnamefont {Schwob}}, \bibinfo {author} {\bibfnamefont {L.}~\bibnamefont {Coolen}}, \bibinfo {author} {\bibfnamefont {A.}~\bibnamefont {Gruzintev}}, \bibinfo {author} {\bibfnamefont {G.}~\bibnamefont {Emel'chenko}}, \bibinfo {author} {\bibfnamefont {V.}~\bibnamefont {Masalov}}, \bibinfo {author} {\bibfnamefont {J.-M.}\ \bibnamefont {Frigerio}},\ and\ \bibinfo {author} {\bibfnamefont {A.}~\bibnamefont {Ma{\^{i}}tre}},\ }\bibinfo {title} {{Manipulating emission of CdTeSe nanocrystals embedded in three-dimensional photonic crystals}},\ \href {https://doi.org/10.1063/1.3129311} {\bibfield  {journal} {\bibinfo  {journal} {J. Appl. Phys.}\ }\textbf {\bibinfo {volume} {105}},\ \bibinfo {pages} {113120} (\bibinfo {year} {2009})}\BibitemShut {NoStop}%
\bibitem [{\citenamefont {Jorgensen}\ \emph {et~al.}(2011)\citenamefont {Jorgensen}, \citenamefont {Galusha},\ and\ \citenamefont {Bartl}}]{Jorgensen2011PRL}%
  \BibitemOpen
  \bibfield  {author} {\bibinfo {author} {\bibfnamefont {M.~R.}\ \bibnamefont {Jorgensen}}, \bibinfo {author} {\bibfnamefont {J.~W.}\ \bibnamefont {Galusha}},\ and\ \bibinfo {author} {\bibfnamefont {M.~H.}\ \bibnamefont {Bartl}},\ }\bibinfo {title} {{Strongly Modified Spontaneous Emission Rates in Diamond-Structured Photonic Crystals}},\ \href {https://doi.org/10.1103/PhysRevLett.107.143902} {\bibfield  {journal} {\bibinfo  {journal} {Phys. Rev. Lett.}\ }\textbf {\bibinfo {volume} {107}},\ \bibinfo {pages} {143902} (\bibinfo {year} {2011})}\BibitemShut {NoStop}%
\bibitem [{\citenamefont {Ning}\ \emph {et~al.}(2012)\citenamefont {Ning}, \citenamefont {Mihi}, \citenamefont {Geddes~III}, \citenamefont {Miyake},\ and\ \citenamefont {Braun}}]{ning2012AdvMater}%
  \BibitemOpen
  \bibfield  {author} {\bibinfo {author} {\bibfnamefont {H.}~\bibnamefont {Ning}}, \bibinfo {author} {\bibfnamefont {A.}~\bibnamefont {Mihi}}, \bibinfo {author} {\bibfnamefont {J.~B.}\ \bibnamefont {Geddes~III}}, \bibinfo {author} {\bibfnamefont {M.}~\bibnamefont {Miyake}},\ and\ \bibinfo {author} {\bibfnamefont {P.~V.}\ \bibnamefont {Braun}},\ }\bibinfo {title} {{Radiative Lifetime Modification of LaF3: Nd Nanoparticles Embedded in 3D Silicon Photonic Crystals}},\ \href@noop {} {\bibfield  {journal} {\bibinfo  {journal} {Adv. Mater.}\ }\textbf {\bibinfo {volume} {24}},\ \bibinfo {pages} {OP153} (\bibinfo {year} {2012})}\BibitemShut {NoStop}%
\bibitem [{\citenamefont {Schulz}\ \emph {et~al.}(2022)\citenamefont {Schulz}, \citenamefont {Harteveld}, \citenamefont {Vancso}, \citenamefont {Huskens}, \citenamefont {Cloetens},\ and\ \citenamefont {Vos}}]{schulz2022acs}%
  \BibitemOpen
  \bibfield  {author} {\bibinfo {author} {\bibfnamefont {A.~S.}\ \bibnamefont {Schulz}}, \bibinfo {author} {\bibfnamefont {C.~A.~M.}\ \bibnamefont {Harteveld}}, \bibinfo {author} {\bibfnamefont {G.~J.}\ \bibnamefont {Vancso}}, \bibinfo {author} {\bibfnamefont {J.}~\bibnamefont {Huskens}}, \bibinfo {author} {\bibfnamefont {P.}~\bibnamefont {Cloetens}},\ and\ \bibinfo {author} {\bibfnamefont {W.~L.}\ \bibnamefont {Vos}},\ }\bibinfo {title} {{Targeted Positioning of Quantum Dots Inside 3D Silicon Photonic Crystals Revealed by Synchrotron X-ray Fluorescence Tomography}},\ \href@noop {} {\bibfield  {journal} {\bibinfo  {journal} {ACS Nano}\ }\textbf {\bibinfo {volume} {16}},\ \bibinfo {pages} {3674} (\bibinfo {year} {2022})}\BibitemShut {NoStop}%
\bibitem [{2()}]{2}%
  \BibitemOpen
  \href@noop {} {}\bibinfo {note} {While the relation between decay rate and lifetime is simple for a single emitter, the relation between distributions of lifetimes and decay rates of an ensemble of emitters that have varying decay rates is complex. Therefore, Eq.~\ref {eq:ldos_ft_to_zetaGamma} cannot directly be applied when using distributions of lifetimes $\zeta (\tau )$ instead of decay rates $\zeta (\Gamma )$.}\BibitemShut {Stop}%
\bibitem [{\citenamefont {Lakowicz}(2006)}]{Lakowicz2006book}%
  \BibitemOpen
  \bibfield  {author} {\bibinfo {author} {\bibfnamefont {J.~R.}\ \bibnamefont {Lakowicz}},\ }\href@noop {} {\emph {\bibinfo {title} {{Principles of Fluorescence Spectroscopy}}}},\ \bibinfo {edition} {3rd}\ ed.\ (\bibinfo  {publisher} {Springer, Berlin},\ \bibinfo {address} {Heidelberg},\ \bibinfo {year} {2006})\BibitemShut {NoStop}%
\bibitem [{\citenamefont {Ho}\ \emph {et~al.}(1994)\citenamefont {Ho}, \citenamefont {Chan}, \citenamefont {Soukoulis}, \citenamefont {Biswas},\ and\ \citenamefont {Sigalas}}]{Ho1994SSC}%
  \BibitemOpen
  \bibfield  {author} {\bibinfo {author} {\bibfnamefont {K.~M.}\ \bibnamefont {Ho}}, \bibinfo {author} {\bibfnamefont {C.~T.}\ \bibnamefont {Chan}}, \bibinfo {author} {\bibfnamefont {C.~M.}\ \bibnamefont {Soukoulis}}, \bibinfo {author} {\bibfnamefont {R.}~\bibnamefont {Biswas}},\ and\ \bibinfo {author} {\bibfnamefont {M.}~\bibnamefont {Sigalas}},\ }\bibinfo {title} {{Photonic Band Gaps in Three Dimensions: {N}}ew layer-{B}y-{L}ayer periodic structures},\ \href {https://doi.org/https://doi.org/10.1016/0038-1098(94)90202-X} {\bibfield  {journal} {\bibinfo  {journal} {Solid State Commun.}\ }\textbf {\bibinfo {volume} {89}},\ \bibinfo {pages} {413} (\bibinfo {year} {1994})}\BibitemShut {NoStop}%
\bibitem [{\citenamefont {Hillebrand}\ \emph {et~al.}(2003)\citenamefont {Hillebrand}, \citenamefont {Senz}, \citenamefont {Hergert},\ and\ \citenamefont {G\"{o}sele}}]{Hillebrand2003JAP}%
  \BibitemOpen
  \bibfield  {author} {\bibinfo {author} {\bibfnamefont {R.}~\bibnamefont {Hillebrand}}, \bibinfo {author} {\bibfnamefont {S.}~\bibnamefont {Senz}}, \bibinfo {author} {\bibfnamefont {W.}~\bibnamefont {Hergert}},\ and\ \bibinfo {author} {\bibfnamefont {U.}~\bibnamefont {G\"{o}sele}},\ }\bibinfo {title} {{Macroporous-silicon-based three-dimensional photonic crystal with a large complete band gap}},\ \href {https://doi.org/10.1063/1.1593796} {\bibfield  {journal} {\bibinfo  {journal} {J. Appl. Phys.}\ }\textbf {\bibinfo {volume} {94}},\ \bibinfo {pages} {2758} (\bibinfo {year} {2003})}\BibitemShut {NoStop}%
\bibitem [{\citenamefont {Vreman}(2025)}]{vreman2025thesis}%
  \BibitemOpen
  \bibfield  {author} {\bibinfo {author} {\bibfnamefont {T.~J.}\ \bibnamefont {Vreman}},\ }\emph {\bibinfo {title} {{Controlling the Reflection and Emission of Light via Photonic Crystals and Quasicrystals}}},\ \href@noop {} {Ph.D. thesis},\ \bibinfo  {school} {University of Twente} (\bibinfo {year} {2025})\BibitemShut {NoStop}%
\bibitem [{3()}]{3}%
  \BibitemOpen
  \href@noop {} {}\bibinfo {note} {Effectively, the inverse woodpile structure also has the fcc structure like the colloidal crystal, albeit with a much more complex crystal basis~\cite {Ashcroft1976book}.}\BibitemShut {Stop}%
\bibitem [{\citenamefont {Oskooi}\ \emph {et~al.}(2010)\citenamefont {Oskooi}, \citenamefont {Roundy}, \citenamefont {Ibanescu}, \citenamefont {Bermel}, \citenamefont {Joannopoulos},\ and\ \citenamefont {Johnson}}]{oskooi2010elsevier}%
  \BibitemOpen
  \bibfield  {author} {\bibinfo {author} {\bibfnamefont {A.~F.}\ \bibnamefont {Oskooi}}, \bibinfo {author} {\bibfnamefont {D.}~\bibnamefont {Roundy}}, \bibinfo {author} {\bibfnamefont {M.}~\bibnamefont {Ibanescu}}, \bibinfo {author} {\bibfnamefont {P.}~\bibnamefont {Bermel}}, \bibinfo {author} {\bibfnamefont {J.~D.}\ \bibnamefont {Joannopoulos}},\ and\ \bibinfo {author} {\bibfnamefont {S.~G.}\ \bibnamefont {Johnson}},\ }\bibinfo {title} {{MEEP: A flexible free-software package for electromagnetic simulations by the FDTD method}},\ \href@noop {} {\bibfield  {journal} {\bibinfo  {journal} {Comput. Phys. Commun.}\ }\textbf {\bibinfo {volume} {181}},\ \bibinfo {pages} {687} (\bibinfo {year} {2010})}\BibitemShut {NoStop}%
\bibitem [{\citenamefont {Johnson}\ and\ \citenamefont {Joannopoulos}(2001)}]{johnson2001OptExp}%
  \BibitemOpen
  \bibfield  {author} {\bibinfo {author} {\bibfnamefont {S.~G.}\ \bibnamefont {Johnson}}\ and\ \bibinfo {author} {\bibfnamefont {J.~D.}\ \bibnamefont {Joannopoulos}},\ }\bibinfo {title} {{Block-iterative frequency-domain methods for Maxwell's equations in a planewave basis}},\ \href@noop {} {\bibfield  {journal} {\bibinfo  {journal} {Opt. Express}\ }\textbf {\bibinfo {volume} {8}},\ \bibinfo {pages} {173} (\bibinfo {year} {2001})}\BibitemShut {NoStop}%
\bibitem [{\citenamefont {Nikolaev}(2006)}]{Nikolaev2006thesis}%
  \BibitemOpen
  \bibfield  {author} {\bibinfo {author} {\bibfnamefont {I.~S.}\ \bibnamefont {Nikolaev}},\ }\emph {\bibinfo {title} {{Spontaneous-emission rates of quantum dots and dyes controlled with photonic crystals}}},\ \href@noop {} {Ph.D. thesis},\ \bibinfo  {school} {University of Twente} (\bibinfo {year} {2006})\BibitemShut {NoStop}%
\bibitem [{\citenamefont {Loudon}(2000)}]{loudon2000book}%
  \BibitemOpen
  \bibfield  {author} {\bibinfo {author} {\bibfnamefont {R.}~\bibnamefont {Loudon}},\ }\href@noop {} {\emph {\bibinfo {title} {{The Quantum Theory of Light}}}}\ (\bibinfo  {publisher} {Oxford University Press},\ \bibinfo {address} {Oxford},\ \bibinfo {year} {2000})\BibitemShut {NoStop}%
\bibitem [{\citenamefont {{Wikipedia contributors}}(2024)}]{wikiWavenumber}%
  \BibitemOpen
  \bibfield  {author} {\bibinfo {author} {\bibnamefont {{Wikipedia contributors}}},\ }\href@noop {} {\bibinfo {title} {{Wavenumber, Wikipedia}}},\ \bibinfo {note} {\url{https://en.wikipedia.org/w/index.php?title=Wavenumber\&oldid=1235416429}}\BibitemShut {NoStop}%
\bibitem [{\citenamefont {Van~Driel}\ \emph {et~al.}(2007)\citenamefont {Van~Driel}, \citenamefont {Nikolaev}, \citenamefont {Vergeer}, \citenamefont {Lodahl}, \citenamefont {Vanmaekelbergh},\ and\ \citenamefont {Vos}}]{vanDriel2007PRB}%
  \BibitemOpen
  \bibfield  {author} {\bibinfo {author} {\bibfnamefont {A.~F.}\ \bibnamefont {Van~Driel}}, \bibinfo {author} {\bibfnamefont {I.~S.}\ \bibnamefont {Nikolaev}}, \bibinfo {author} {\bibfnamefont {P.}~\bibnamefont {Vergeer}}, \bibinfo {author} {\bibfnamefont {P.}~\bibnamefont {Lodahl}}, \bibinfo {author} {\bibfnamefont {D.}~\bibnamefont {Vanmaekelbergh}},\ and\ \bibinfo {author} {\bibfnamefont {W.~L.}\ \bibnamefont {Vos}},\ }\bibinfo {title} {{Statistical analysis of time-resolved emission from ensembles of semiconductor quantum dots: Interpretation of exponential decay models}},\ \href@noop {} {\bibfield  {journal} {\bibinfo  {journal} {Phys. Rev. B}\ }\textbf {\bibinfo {volume} {75}},\ \bibinfo {pages} {035329} (\bibinfo {year} {2007})}\BibitemShut {NoStop}%
\bibitem [{\citenamefont {Ashcroft}\ and\ \citenamefont {Mermin}(1976)}]{Ashcroft1976book}%
  \BibitemOpen
  \bibfield  {author} {\bibinfo {author} {\bibfnamefont {N.~W.}\ \bibnamefont {Ashcroft}}\ and\ \bibinfo {author} {\bibfnamefont {N.~D.}\ \bibnamefont {Mermin}},\ }\href@noop {} {\emph {\bibinfo {title} {{Solid State Physics}}}}\ (\bibinfo  {publisher} {{Holt, Rinehart and Winston}},\ \bibinfo {address} {{New York}},\ \bibinfo {year} {1976})\BibitemShut {NoStop}%
\end{thebibliography}%

\end{document}